\documentclass[%
    aps,
    prd,
    letterpaper,
    superscriptaddress,
    preprintnumbers,
    floatfix,
    twocolumn,
    nofootinbib,
    showpacs,
    hyphens
]{revtex4-2}  

\usepackage[normalem]{ulem}
\usepackage{subfigure}
\usepackage{amsfonts}
\usepackage{graphicx}
\usepackage{dcolumn}
\usepackage{bm}
\usepackage{hyperref}
\usepackage{xcolor}
\usepackage{amsmath,amsfonts,amssymb,amsthm}
\newcommand{\ket}[1]{\left| #1 \right>} 
\newcommand{\bra}[1]{\left< #1 \right|} 
\newcommand{\braket}[2]{\left< #1 \vphantom{#2} \right|
 \hspace{-2pt} \left. #2 \vphantom{#1} \right>} 
\newcommand{\mbraket}[3]{\left< #1 \vphantom{#2#3} \right|
 #2 \left| #3 \vphantom{#1#2} \right>} 

\usepackage{slashed}

\newcommand{\Em}[0]{E_0^{(m)}}
\newcommand{\Tm}[0]{T^{(m)}}
\newcommand{\Tbm}[0]{T^{(b,m)}}
\newcommand{\CWm}[0]{\widetilde{T}^{(m)}}
\newcommand{\CWbm}[0]{\widetilde{T}^{(b,m)}}

\usepackage{duckuments}

\begin{document}
\title{Lanczos, the transfer matrix, and the signal-to-noise problem}
\author{Michael L. Wagman}
\affiliation{Fermi National Accelerator Laboratory, Batavia, IL 60510, USA}  

\preprint{FERMILAB-PUB-24-0320-T}

\renewcommand{\vec}[1]{\boldsymbol{#1}}
\DeclareRobustCommand{\Eq}[1]{Eq.~\eqref{eq:#1}}
\DeclareRobustCommand{\Eqs}[2]{Eqs.~\eqref{eq:#1} and \eqref{eq:#2}}
\DeclareRobustCommand{\fig}[1]{Fig.~\ref{fig:#1}}
\DeclareRobustCommand{\figs}[2]{Figs.~\ref{fig:#1} and \ref{fig:#2}}
\DeclareRobustCommand{\app}[1]{App.~\ref{app:#1}}
\DeclareRobustCommand{\sec}[1]{Sec.~\ref{sec:#1}}
\DeclareRobustCommand{\secs}[2]{Secs.~\ref{sec:#1} and \ref{sec:#2}}
\DeclareRobustCommand{\tbl}[1]{Table~\ref{tbl:#1}}
\DeclareRobustCommand{\refcite}[1]{Ref.~\cite{#1}}
\DeclareRobustCommand{\refcites}[1]{Refs.~\cite{#1}}

\begin{abstract}
This work introduces a method for determining the energy spectrum of lattice quantum chromodynamics (LQCD) by applying the Lanczos algorithm to the transfer matrix and using a bootstrap generalization of the Cullum-Willoughby method to filter out spurious eigenvalues.
  Proof-of-principle analyses of the simple harmonic oscillator and the LQCD proton mass demonstrate that this method provides faster ground-state convergence than the ``effective mass,'' which is related to the power-iteration algorithm.
Lanczos provides more accurate energy estimates than multi-state fits to correlation functions with small imaginary times while achieving comparable statistical precision.
Two-sided error bounds are computed for Lanczos results and guarantee that excited-state effects cannot shift Lanczos results far outside their statistical uncertainties.
\end{abstract}

\maketitle

Determining the energy spectrum of quantum chromodynamics (QCD) leads to hadron mass predictions and is an essential first step towards predicting hadron and nuclear structure, scattering amplitudes, and inputs to new physics searches.
Spectroscopy, the determination of ground- and excited-state energies, is therefore a central aspect of lattice QCD (LQCD) calculations.
Accurately disentangling ground- and excited-state effects is challenging for systems with small energy gaps and leads to significant and hard-to-quantify uncertainties in state-of-the-art LQCD calculations for example of nucleon axial form factors~\cite{Jang:2019vkm,RQCD:2019jai,Alexandrou:2020okk,Park:2021ypf,Djukanovic:2022wru,Jang:2023zts,Alexandrou:2023qbg,Gupta:2024krt} and baryon-baryon scattering~\cite{Francis:2018qch,Horz:2020zvv,Amarasinghe:2021lqa,Green:2021qol,Detmold:2024iwz}. 

Spectroscopy calculations typically rely on analyzing the asymptotic decay rate of Euclidean correlation functions, which is set by the lowest-energy state not orthogonal to those created by the relevant operators.
However, the signal-to-noise ratios (SNRs) of Euclidean correlation functions also decay exponentially in Euclidean time with a rate predicted by Parisi and Lepage~\cite{Parisi:1983ae,Lepage:1989hd}.
This so-called SNR problem is a major challenge for proton and other baryon correlation functions and becomes exponentially more severe for nuclei~\cite{Beane:2009gs,Beane:2009py,Davoudi:2020ngi}.

The SNR problem has motivated the development of methods to maximize the information that can be extracted from relatively precise correlation functions with small imaginary times. Techniques based on solving a generalized eigenvalue problem (GEVP) constructed from a symmetric correlation-function matrix provide both faster convergence and rigorous variational upper bounds on energies~\cite{Fox:1981xz,Michael:1982gb,Luscher:1990ck,Blossier:2009kd,Fleming:2023zml}, and their application has become standard in LQCD spectroscopy for multi-hadron systems with small energy gaps~\cite{Briceno:2017max,Bulava:2022ovd,Hanlon:2024fjd}.
Prony's method for signal processing has also been applied to LQCD and shown to improve ground-state convergence~\cite{Fleming:2004hs,Lin:2007iq,Fleming:2009wb,Beane:2009kya,Fischer:2020bgv}.

This work proposes a new approach to LQCD spectroscopy using the Lanczos algorithm~\cite{Lanczos:1950zz} to compute eigenvalues of the transfer matrix.
The Lanczos algorithm has been widely applied for decades to computational linear algebra~\cite{Parlett,Golub:1989,Parlett:1995,Meurant:2006,Saad:2011,Golub:2013,str-5}, quantum Monte Carlo calculations~\cite{Caffarel:1991,Sorella:2001,Becca:2017}, and analysis of Dirac matrices in LQCD~\cite{Barbour:1983bk,Barbour:1984ud,Barbour:1985dq,Kalkreuter:1995vg,Kennedy:1998cu,Clark:2017wom,Jeong:2022hpg}.
However, direct application of the Lanczos algorithm to the LQCD transfer matrix is challenging because the transfer matrix is infinite-dimensional~\cite{Kogut:1974ag} and not directly constructed in path-integral Monte Carlo calculations.

Here, I show that the Lanczos algorithm can be applied to the LQCD transfer matrix using recursive formulae whose inputs are simply Euclidean correlation functions.
The Kaniel-Paige-Saad (KPS) bound~\cite{Kaniel:1966,Paige:1971,Saad:1980} implies that Lanczos energies converge exponentially faster than standard estimators near the continuum limit.
Lanczos approximation errors can be estimated directly and provide two-sided error bounds, improving upon the one-sided variational bounds provided by GEVP solutions.

Proof-of-principle results are shown below for a quantum simple harmonic oscillator (SHO) and the proton mass in LQCD with close-to-physical quark masses.
The variance of Lanczos estimators is found to approach a constant for large iteration counts, in contrast to algebraic estimators based on ``effective masses'' whose variance grows exponentially in accordance with Parisi-Lepage scaling.
An explanation for this behavior is proposed based on the identification of Lanczos as a Krylov-space projection method and anlysis of projection operator SNRs inspired by Della Morte and Giusti~\cite{DellaMorte:2007zz,DellaMorte:2008jd,DellaMorte:2010yp}.
Of course, extending the maximum time included in a multi-state fit offers the same improvement.
Unlike fits, however, Lanczos results are purely algebraic functions of LQCD ``data'' and do not require any statistical inference, cuts on what subset of data is included, Bayesian priors, or covariance-matrix estimation.
In examples where only small ranges of data are available, Lanczos results simultaneously achieve similar variance and less bias than multi-state fit results.

{\it \bf Method:}
Lattice field theories do not have a continuous time-translation symmetry to be used to directly define a Hamiltonian operator $H$.
The presence of discrete time-translation symmetry, $t \rightarrow t+a$ where $a$ is the lattice spacing, does allow a transfer matrix $T = e^{- a H}$ to be defined that acts as an imaginary-time evolution operator.
The transfer matrix for LQCD with the Wilson gauge and fermion actions has been explicitly constructed as a Hilbert-space operator and shown to be Hermitian and positive-definite~\cite{Luscher:1976ms}.
A transfer matrix can also be constructed for theories with improved actions that acts as a positive-definite operator for low-energy states~\cite{Luscher:1984is}.

Euclidean correlation functions for a theory with temporal extent (inverse temperature) $\beta$ and ``interpolating operators'' $\psi$ and $\psi^\dagger$ are matrix elements of powers of $T$,
\begin{equation}
\begin{split}
    C(t) &\equiv \left< \psi(t) \psi^\dagger(0) \right> 
    = \mbraket{\psi}{T^{t/a}}{\psi} + \ldots,
    \end{split}
\end{equation}
where $\ket{\psi} \equiv \psi^\dagger \ket{0}$ with $\ket{0}$ the vacuum state and ... denotes thermal effects detailed in the Supplemental Material, which also includes Refs.~\cite{Hackett:2023nkr,Abbott:2024vhj}.
Inserting complete sets of transfer-matrix eigenstates $\sum_n \ket{n}\bra{n}$ leads to the spectral representation
\begin{equation}
\begin{split}
    C(t) 
    &= \sum_n \lambda_n^t \braket{\psi}{n}\braket{n} {\psi} = \sum_{n=0}^{\infty} |Z_n|^2 e^{-E_n t},
    \label{eq:spec}
    \end{split}
\end{equation}
where $Z_n \equiv \braket{n}{\psi}$ are overlap factors. Energies are defined by $E_n \equiv -(1/a) \ln \lambda_n$ in terms of transfer matrix eigenvalues 
$\lambda_n \equiv \mbraket{n}{T}{n}$ in the sector with quantum numbers of $\psi^\dagger \ket{0}$ ordered such that $\lambda_0 \geq \lambda_1 \geq \ldots$.
For large $t$, correlation functions are dominated by contributions from the lowest-energy state that is not orthogonal to $\ket{\psi}$.
The ``effective mass'' is an algebraic estimator for the ground-state energy
\begin{equation}
    E(t) \equiv -\frac{1}{a} \ln\left[ \frac{C(t)}{C(t-a)} \right] 
    = E_0 + O\left( e^{-(t/a) \delta } \right),
\end{equation}
where $\delta \equiv a(E_1 - E_0)$ is the excitation gap. 

Estimating the ground-state energy using the effective mass can be viewed as an application of the power-iteration method~\cite{VonMises:1929} to the transfer matrix as follows.
This method defines $\ket{b_k} \equiv T \ket{b_{k-1}} / ||T \ket{b_{k-1}}||$ with $\ket{b_1} = \ket{\psi} / \sqrt{\braket{\psi}{\psi}}$ 
which leads to $\ket{b_k} \propto T^{k-1} \ket{\psi}$.
For Hermitian $T$, the Hilbert-space norm satisfies $||T \ket{\psi}|| = \mbraket{\psi}{T^2}{\psi} = C(2a)$.
The largest eigenvalue of $T$ is then approximated as
\begin{equation}
  \mu_k \equiv \frac{\mbraket{b_k}{ T }{b_k}}{ \braket{b_k}{b_k} } = \frac{ \mbraket{\psi}{T^{2k-1}}{\psi}}{ \mbraket{\psi}{T^{2(k-1)}}{\psi}} \\
    = \frac{C((2k-1)a)}{C((2k-2) a)}.
\end{equation}
Taking the log gives the power-iteration energy estimate after $k$ iterations: $-(1/a) \ln \mu_k$.
This estimate is identical to the effective mass with $t/a = 2k-1$.

The Lanczos algorithm~\cite{Lanczos:1950zz} is widely appreciated to be superior to the power-iteration method for the task of approximating the largest eigenvalue of a matrix~\cite{Parlett,Parlett:1982,Golub:1989,Kuczynski:1992,Parlett:1995,Meurant:2006,Kuijlaars:2000,Saad:2011,Golub:2013,Garza-Vargas:2020}.
Both methods involve the Krylov space
\begin{equation}
    \mathcal{K}^{(m)} = \text{span}\{ \ket{\psi}, T\ket{\psi}, \ldots, T^m \ket{\psi} \},
\end{equation}
and the approximation of eigenvectors as elements of $\mathcal{K}^{(m)}$.
However, the Lanczos algorithm enables the explicit diagonalization of a Krylov-space approximation to $T$ that leads to faster convergence as discussed below.
Lanczos vectors are defined by the three-term recurrence
\begin{equation}
  T\ket{v_j} = \alpha_j\ket{v_j} + \beta_{j}\ket{v_{j-1}} + \beta_{j+1} \ket{v_{j+1}}. \label{eq:Lanczos}
\end{equation}
This guarantees that $\braket{v_i}{v_j} = \delta_{ij}$ and therefore $\{ \ket{v_1},\ldots,\ket{v_m} \}$ provides an orthonormal basis for $\mathcal{K}^{(m)}$.
The matrix elements of $T$ in this basis, $\Tm_{ij} \equiv \mbraket{v_i}{T}{v_j}$, form a tridiagonal matrix by Eq.~\eqref{eq:Lanczos}.
Its eigenvalues $\lambda_n^{(m)}$, called ``Ritz values,'' provide optimal Krylov-space approximations to eigenvalues of $T$~\cite{Parlett}.

The LQCD transfer matrix is an infinite-dimensional integral operator that cannot be explicitly represented numerically without some form of Hilbert-space truncation.
However, the elements of $\Tm_{ij}$ are scalars constructed from inner products involving $\ket{\psi}$ and $T$.
The path integral definitions of $\Tm_{ij}$ can be related to those for Euclidean correlation functions $C(ka)$ by defining normalized states $\ket{v_1} = \ket{\psi} / \sqrt{\braket{\psi}{\psi}}$, as in the first step of the power-iteration method, and matrix elements $A_j^{(p)} \equiv \mbraket{v_j}{T^p}{v_j}$ and $B_j^{(p)} \equiv \mbraket{v_j}{T^p}{v_{j-1}}$.
For $j=1$, $A_1^{(p)} = C(pa)/C(0)$; by definition $B_1^{(p)} = 0$.
Recursion relations
\begin{equation} \label{eq:algo_diag}
\begin{split}
    A_{j+1}^{(p)} &= \frac{1}{\beta_{j+1}^2} \left[ 
    A_j^{(p+2)} 
    + \alpha_j^2 A_j^{(p)}   
    + \beta_j^2 A_{j-1}^{(p)} - 2\alpha_j A_j^{(p+1)}  \right. \\
    &\hspace{40pt} \left. 
    + 2\alpha_j \beta_j B_j^{(p)}
    - 2\beta_j B_j^{(p+1)}   \right],
    \end{split}
\end{equation}
and
\begin{equation} \label{eq:algo_off_diag}
\begin{split}
    B_{j+1}^{(p)} &= \frac{1}{\beta_{j+1}} \left[ 
    A_j^{(p+1)} 
    - \alpha_j A_j^{(p)}
    - \beta_j B_j^{(p)} \right],
    \end{split}
\end{equation}
provide matrix elements for $j>1$,
where $p \in \{1, \ldots 2(m-j)+1 \}$, $\alpha_j = A_j^{(1)}$, $\beta_1 = B_1^{(p)} = 0$, and $\beta_{j+1} = \sqrt{A_j^{(2)} - \alpha_j^2 - \beta_j^2} = B_{j+1}^{(1)}$. 
Similar recursions are discussed for Monte Carlo thermalization times in Ref.~\cite{DeMeo:1998}.

These recursions provide the non-zero matrix elements $\Tm_{ii} = A_i^{(1)} = \alpha_i$ and $\Tm_{i(i+1)} = \Tm_{(i+1)i} = B_{i+1}^{(1)} = \beta_{i+1}$.
The $m\times m$ matrix $\Tm_{ij}$ can be diagonalized straightforwardly to provide $\lambda_k^{(m)}$ and $E_k^{(m)} \equiv -(1/a) \ln \lambda_k^{(m)}$ with $k \in \{1,\ldots,m\}$.
The $m=1$ case simplifies to $T^{(1)}_{11} = \alpha_1 = C(a)/C(0)$ and therefore $E_0^{(1)} = E(a)$.

The recursion relations shown here are applicable to any $C(t)$ admitting a spectral representation; see the Supplemental Material for a proof.
They further apply to stochastic estimates of $C(t)$ if $\Tm_{ij} \in \mathbb{C}$ is permitted\footnote{Non-spurious eigenvalues with $\beta_j^2 \approx 0$ indicate that a Ritz value has converged to an eigenvalue of $T$, see Refs.~\cite{Paige:1971,Parlett:1979,Parlett,Parlett:1995}.  After convergence has been achieved, statistical fluctuations can easily lead to $\beta_j^2 < 0$ and therefore complex $T^{(m)}_{ij}$.}
  and they are interpreted as applying oblique Lanczos~\cite{Saad:1982,Parlett:1985,Nachtigal:1993} with a non-Hermitian transfer matrix $T^{(m)}$ that exactly describes $C(t)$ at finite statistics, as proven in Ref.~\cite{Hackett:2024xnx}.
Unless stated otherwise, theoretical results below assume that $T^{(m)}$ is Hermitian and therefore only apply 
 exactly at infinite statistics and stochastically at finite statistics

{\it \bf Convergence:} 
Cauchy's interlacing theorem guarantees that $\lambda_k^{(m)}$ provides a lower bound on $\lambda_k$~\cite{Wilkinson:1965,Paige:1971,Parlett}.
Therefore $E_k^{(m)}$ provides a variational upper bound on $E_k$.
The KPS bound~\cite{Kaniel:1966,Paige:1971,Saad:1980} further constrains the approximation errors of ground-state Ritz values:
\begin{equation}\label{eq:KPS0} \begin{split}
    \frac{\lambda_0 - \lambda_0^{(m)}}{\lambda_0} &\leq  \left[ \frac{\tan \arccos Z_0}{T_{m-1}(2 e^{\delta} - 1)} \right]^2,
    \end{split}
\end{equation}
where the $T_k(x)$ are Chebyshev polynomials of the first kind, $T_k(\cos x) = \cos(k x)$.
For large $k$, $T_{k}(x) \approx \frac{1}{2}(x + \sqrt{x^2 - 1})^{k}$, and the KPS bound simplifies 
to
\begin{equation}
    \frac{ \lambda_0 - \lambda_0^{(m)}}{\lambda_0 } \lesssim  \frac{4 (1 - Z_0^2)}{Z_0^2} \times \begin{cases} e^{-2 (m-1) \delta }    &  \delta \gg 1 \\
    e^{-4(m-1)\sqrt{\delta}} &  \delta \ll 1 
    \end{cases} .
\end{equation}
For large $\delta$, this resembles the $O( e^{-(t/a) \delta })$ asymptotic error of the effective mass (power-iteration method) with $t/a \propto 2m$.
For small $\delta$, Lanczos converges exponentially faster than the power-iteration method because $\sqrt{\delta} \gg \delta$.
The latter region is the relevant one near the continuum limit where $\delta = a(E_1 - E_0) \ll 1$.

\begin{figure}[t!]
                \centering
                \includegraphics[width=0.48\textwidth]{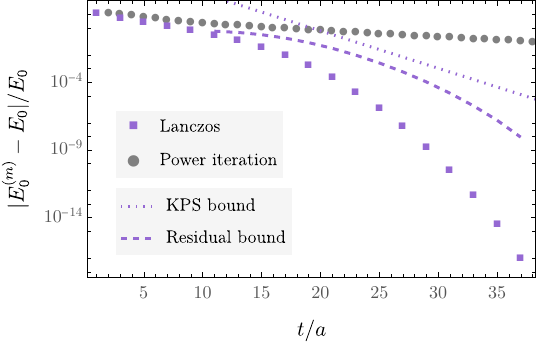}
            \caption{\label{fig:mock_correlator}
                Relative errors of Lanczos and power-iteration (effective mass) applied to $\{C(0),\ldots,C(t a)\}$ versus $t/a = 2m-1$ for the 20-state mock data described in the main text (including the trivial extension of power-iteration to half-integer $m$). 
                }
                The Kaniel-Paige-Saad bound, Eq.~\eqref{eq:KPS0}, is shown as a dotted line. The residual bound, Eq.~\eqref{eq:Rbound}, is shown as a dashed line for $m\geq 6$ ($\Em$ is closer to $E_n$ with $n>0$ for smaller $m$).
\end{figure}

Fig.~\ref{fig:mock_correlator} compares the practical convergence of $E_0^{(m)}$ with that of $E(t)$ for a 20-state correlation-function model defined by Eq.~\eqref{eq:spec} with $Z_n = (n+1)$ and $a E_n = 0.1 (n+1)$ for $n=0,\ldots,19$. 
The Ritz values exactly reproduce the complete energy spectrum after 20 iterations and for the ground state are 5 orders of magnitude more accurate than the standard effective mass after 12 iterations.

Remarkably, it is also possible to bound the difference between $E_k^{(m)}$ and an exact energy in terms of directly calculable quantities~\cite{Parlett,Parlett:1995}.
The approximation error of $T$ in the Lanczos basis is the residual norm $\beta_{m+1}$ by Eq.~\eqref{eq:Lanczos}, and the change of basis relating this to the Ritz basis is provided by the eigenvectors $\omega_n^{(m)}$ of $\Tm_{ij}$.
The residual bound states that for any Ritz value $\lambda_k^{(m)}$ there exists an eigenvalue $\lambda$ of $T$ satisfying 
\begin{equation}\label{eq:Rbound}
  \min_{\lambda \in \{\lambda_n\}} |\lambda_k^{(m)} - \lambda |^2 \leq B_k^{(m)} \equiv |\beta_{m+1}|^2 \, |\omega_{mk}^{(m)}|^2\, V_k^{(m)},
\end{equation}
where $\omega_{jk}^{(m)}$ denotes the $j$-th component of $\omega_k^{(m)}$. 
Here, $V_k^{(m)} \equiv \bigl\langle v_{m+1} | v_{m+1} \bigr\rangle / \bigl\langle y_k^{(m)} | y_k^{(m)} \bigr\rangle$ involves the Ritz vectors $\bigl|y_n^{(m)} \bigr> \equiv \sum_j \bigl| v_j^{(m)} \bigr> \omega_{jn}^{(m)} $ and is equal to unity when $\Tm_{ij}$ is Hermitian. 
Importantly, the residual bound does not assume Hermiticity of $T^{(m)}$.
Its evaluation in the context of oblique Lanczos is detailed in the Supplemental Material.
Note that it is not guaranteed that $\lambda_k^{(m)}$ is close to $\lambda_k$ --- eigenvectors that have sufficiently small overlap with $\ket{\psi}$ can be ``missed'' by Lanczos~\cite{Parlett,Parlett:1990,Kuijlaars:2000} --- but a two-sided systematic uncertainty interval in which an eigenvalue of $T$ is guaranteed to exist can be computed for each $\lambda_k^{(m)}$.
This is a significant advantage over standard methods, which can also miss energy eigenstates in practice~\cite{Dudek:2012xn,Lang:2012db,Wilson:2015dqa,Amarasinghe:2021lqa} and for finite $t$ only have one-sided systematic uncertainty bounds arising from the variational principle.

It is noteworthy that an improved estimator based on Prony's method~\cite{Fleming:2004hs,Lin:2007iq,Fleming:2009wb,Beane:2009kya,Fischer:2020bgv} gives results that are numerically identical to the $\lambda_k^{(m)}$ when applied to the $2m$ correlation-function values $\{C(0),\ldots,C((2m-1)a)\}$. 
After this work was completed, this and further coincidences between Lanczos and other methods were explored~\cite{Ostmeyer:2024qgu,Chakraborty:2024scw} and placed in the context of a larger equivalence class~\cite{Abbott:2025yhm}. 
Here, I focus on important properties of the $\lambda_k^{(m)}$---including the KPS and residual bounds discussed above, as well as methods for filtering between physical and spurious Ritz values in applications to noisy data discussed below---that can be derived from their Lanczos definitions and are not obvious from the perspective of Prony's method.\footnote{LQCD applications of Prony's method have used fixed $m$ in the range $m \leq 4$~\cite{Lin:2007iq,Fleming:2009wb,Beane:2009kya,Fischer:2020bgv} and noted that unphysical solutions arising from noise become increasingly common as $m$ is increased~\cite{Fleming:2009wb}.}

{\it \bf Numerical Stability:} 
Numerical artifacts in Lanczos arising from finite-precision arithmetic have been studied in detail and result in ``spurious eigenvalues'' that do not converge towards definite values as well as multiple copies of genuine eigenvalues~\cite{Paige:1971,Parlett,Cullum:1981,Cullum:1985}.
A simple yet effective~\cite{Kalkreuter:1995vg,Elsner:1999} method for filtering spurious eigenvalues was introduced by Cullum and Willoughby~\cite{Cullum:1981,Cullum:1985}.
This method is based on the non-trivial realization that large differences between Ritz values and $T$ eigenvalues can only arise for a non-degenerate eigenvalue when $\lambda_n^{(m)}$ is also an eigenvalue of the matrix $\CWm_{ij}$ defined by deleting the first row and column of $\Tm_{ij}$~\cite{Cullum:1981,Cullum:1985}.

The threshold for when the distance $d_k^{(m)}$ between an eigenvalue of $\Tm_{ij}$ and the nearest eigenvalue of $\CWm_{ij}$ should be considered zero --- and thus the eigenvalue spurious --- is obvious in fixed-precision Lanczos applications without statistical noise~\cite{Cullum:1981,Cullum:1985} but for Monte Carlo results it depends on statistical precision.
Since spurious eigenvalues appear because of statistical fluctuations and do not converge to definite values, the distribution of Ritz values $\lambda_k^{(b,m)}$ obtained using bootstrap resampling~\cite{Efron:1982,Davison:1997,Young:2014,gvar} with $b \in \{1,\ldots,N_{\rm boot}\}$ is useful for identifying 
 a threshold $\varepsilon_{CW}$ for which eigenvalues with $d_k^{(m)} < \varepsilon_{CW}$ are deemed spurious as described below.

This construction does an imperfect job of filtering between spurious and non-spurious eigenvalues in the presence of noise.
Using outlier-robust bootstrap statistics, in particular the bootstrap median and its empirical bootstrap confidence intervals, can mitigate the statistical noise that results from misidentification.

Lanczos estimates for $E_0$ are therefore obtained by:
\begin{enumerate}
  \item Compute the Ritz values $\lambda_k^{(b,m)}$ for $N_{\rm boot}$ bootstrap samples using Lanczos recursions, Eq.~\eqref{eq:algo_diag}-\eqref{eq:algo_off_diag}. If $|\text{arg}\lambda_k^{(b,m)}| > \varepsilon_{\rm float} \sim 10^{-12}$, discard as spurious.
  \item Place the Cullum-Willoughby threshold, $\varepsilon_{CW}$, below all $\ln d^{(b,m)}$ histogram bins with $O(N_{\rm boot} m)$ counts.\footnote{A precise recipe for defining $\varepsilon_{CW}$ in terms of three hyperparameters is given in the Supplemental Material; examples here use $\Delta = 4$, $K_{CW} = 3$, and $F_{CW} = 10$.} If $d_k^{(b,m)} > \varepsilon_{CW}$, discard as spurious. 
  \item $E_0 = -\ln \text{median}_b[ \lambda_0^{(b,m_{\rm max})} ]$, where $\lambda_0^{(b,m)}$ is the largest non-spurious $\lambda_k^{(b,m)}$ less than one and $m_{\rm max} = (\beta-1)/2$ is the largest $m$ where $B_k^{(b,m)}$ is computable.
  \item To compute uncertainties, use a nested resampling procedure wherein, for each ``outer'' bootstrap ensemble, the median is taken over estimates computed on $N_{\rm boot}$ ``inner'' bootstrap ensembles drawn from the outer one. Empirical confidence intervals of the outer bootstrap results provide (idiomatic) bootstrap uncertainties for $E_0$.
\end{enumerate}
Lanczos estimates for $B_0$ are computed from $B_k^{(b,m_{\rm max})}$ using an identical nested bootstrap procedure.

{\it \bf Simple harmonic oscillator:}
Complex scalar field theory in $(0+1)D$ with periodic boundary conditions, $\varphi(\beta)=\varphi(0)$, and action
\begin{equation}
\begin{split}
    S &= a\sum_{t=0}^{\beta - 1} \left\lbrace \varphi(t)^* \left[ \varphi(t + a) - 2\varphi(t) + \varphi(t - a) \right]\frac{1}{a^2} \right. \\
    &\left. \hspace{50pt}+ (aM)^2 |\varphi(t)|^2 \vphantom{\frac{1}{a^2}} \right\rbrace,
    \end{split}
\end{equation}
describes a system of non-interacting bosons with mass $aM$ at temperature $1/\beta$. For $\beta \rightarrow \infty$ it is equivalent to a pair of simple harmonic oscillators.
The theory has a $U(1)$ symmetry $\varphi \rightarrow e^{i \theta} \varphi$ whose conserved charge $Q$ corresponds to $\varphi$ particle number.
Correlation functions for systems with charge $Q$ have ground-state energies $Q m$.
Since $|\varphi|^2$ has $Q=0$, the variances of charge $Q$ correlation functions are asymptotically constant.
This gives calculations of $Q=1$ correlation functions an exponentially severe SNR problem
that is similar to the SNR problem facing baryons in LQCD~\cite{Wagman:2016bam,Detmold:2018eqd}.

\begin{figure}[t!]
                \centering
                \includegraphics[width=0.48\textwidth]{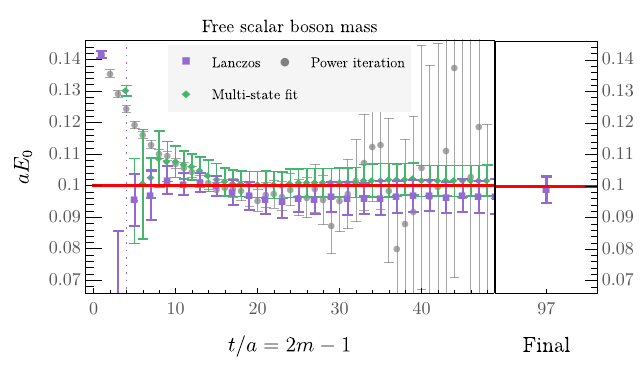}
            \caption{\label{fig:SHO_correlator}
 Lanczos results for $E_0$ are compared with power-iteration (effective mass) results after $m$ iterations as a function of the maximum $t/a = 2m-1$ for which $C(t)$ is included. Multi-state fit results are shown as functions of the maximum $t$ for which $C(t)$ is included. The exact result is a red line. }
\end{figure}

A correlation function with $Q=1$ that has overlap with both ground and excited states can be defined as
\begin{equation}
    C_{\varphi}(t) \equiv \left< \varphi(t)|\varphi(t)|^{3/2} \varphi(0)^* |\varphi(0)|^{3/2} \right>.
\end{equation}
Results for $E(t)$ and $E_0^{(m)}$ are shown in Fig.~\ref{fig:SHO_correlator} for an ensemble of $10^4$ field configurations with $aM = 0.1$ and $\beta/a=100$ where $C_{\varphi}(t)$ is averaged over all translations of the origin. 
Blocking analysis shows that averaging over $N_{\rm block} = 10$ configurations is sufficient to provide negligible autocorrelations.
Uncertainties are calculated using (nested) bootstrap confidence intervals with $N_{\rm boot} = 200$.

Lanczos results converge within uncertainties to the exact result $aE_0 \approx 0.09996$ after $m=3$ iterations, corresponding to $t/a=5$, while $E(t)$ reaches similar convergence for $t/a \sim 10$.
Further, $E_0^{(m)}$ converges to the true ground-state energy even in the presence of thermal states, as proven in the Supplemental Material, and Lanczos results remain consistent with $aM$ for $t > \beta/2$.
The Lanczos SNR also converges to a constant value for large $m$.
This is very different from the power-iteration (effective-mass) SNR, which decreases exponentially according to Parisi-Lepage scaling.
In this way Lanczos SNR scaling offers the same improved SNR scaling of multi-state fit results as a function of the maximum time $t_{\rm max}$ included in fits.

Applying Lanczos for the full $m_{\rm max} = (\beta - 1)/2 = 49$ iteration count available gives $E_0^{(m_{\rm max})} = 0.0987(40)$.
The two-sided window provided by the residual bound is $[ 0.065(12), 0.128(10)]$, where $-\ln(\lambda_k^{(m)} \pm B_k^{(m)})$ is computed for the $m$ where $B_k^{(m)}$ is minimized and uncertainties correspond to bootstrap empirical confidence intervals for $-\ln( \lambda_k^{(b,m)} \pm \sqrt{ B_k^{(b,m)} })$. 
This window is an order of magnitude larger than the statistical uncertainty of $E_0^{(m_{\rm max})}$ but nonetheless provides a meaningful constraint on the possible size of excited-state effects.
Since the minimum $B_k^{(m)} = 0.00083(69)$ is not clearly resolved from zero, the scaling of the size of this two-sided residual-bound window with statistical ensemble size $N$ is non-trivial and important to explore in future work.

This can be compared to standard multi-state fits to $C_{\varphi}(t)$ using the methodology for averaging over $t_{\rm min}$ and choosing $N_{\rm states}$ detailed in Refs.~\cite{NPLQCD:2020ozd,Amarasinghe:2021lqa}: a weighted average of multi-state fits with all $t_{\rm min}/a \geq 1$ gives $0.1016(52)$.
Lanczos and this weighted average both agree with the exact result within $1\sigma$; the Lanczos uncertainties are $\approx 30\%$ smaller.
Here, as in many LQCD applications, the sample covariance matrix is ill-conditioned; I regulate it using linear shrinkage~\cite{Ledoit:2004} as in Ref.~\cite{Rinaldi:2019thf}.
Several other choices are required to precisely define the fitting scheme: $N_{\rm states}$ for each fit range is chosen using the Akaike Information Criterion (AIC)~\cite{AkaikeAIC} using $\rm{tol}_{AIC} = -0.5$~\cite{NPLQCD:2020ozd}, only fit ranges with at least $t_{\rm plateau} = 6$~\cite{NPLQCD:2020ozd} points are included, and numerical checks on optimizer convergence are implemented as in Ref.~\cite{NPLQCD:2020ozd}.
Both Lanczos and multi-state fit results converge to the exact ground-state energy within $1\sigma$ for $t_{\rm max}/a \sim 2m-1 \sim 5$, as long as one-state fits to short time ranges are excluded ($t_{\rm plateau} \geq 5$).

Fits with small $t_{\rm min}$ can be inaccurate even when $\chi^2/\text{dof} < 1$: one-state fits with $t_{\rm min} = 7$ and $t_{\rm max} = 35$ (chosen via SNR cutoff~\cite{NPLQCD:2020ozd}) give $0.1079(17)$, a $4\sigma$ discrepancy, with $\chi^2/\text{dof} = 0.74$.
This underscores the importance of the residual bound, which is rigorously valid even for small $m$.
As an extreme case, $E_0^{(1)} = 0.1416(12)$ has clearly not converged within uncertainties, but $B_0^{(1)} = 0.0045(3)$ still provides a two-sided window $[0.0672(26),0.2219(34)]$ that includes the true ground-state energy.

{\it \bf Lattice QCD:}
Correlation functions for baryons and nuclei in LQCD have exponential SNR problems that can make it difficult to isolate ground states.
To test the accuracy and precision of Lanczos methods for baryons, I computed proton correlation functions $C_p(t) = \left< p(t) \overline{p}(0) \right>$ using point-like proton interpolating operators $p(x) = u^T(x) C \gamma_5 d(x) (1 + \gamma_4) d(x)$ projected to zero spatial momentum for 64 sources on a single timeslice of 76 gauge-field configurations with $L/a = 48$, $\beta/a = 96$, and $N_f = 2+1$ dynamical quarks with light quark masses corresponding to $m_\pi \approx 170$ MeV and $a \approx 0.091(1)$ fm~\cite{Yoon:2016jzj,Mondal:2020ela}.
The action corresponds to the L{\"u}scher-Weisz gauge action~\cite{Luscher:1984xn} and clover-improved~\cite{Sheikholeslami:1985ij} Wilson fermion action with one step of stout smearing~\cite{Morningstar:2003gk}; a Hermitian transfer matrix representation is therefore only valid for low-energy states~\cite{Luscher:1984is}.
Results for $C(t)$ and $C(-t)$ are averaged.
Autocorrelations are found to be negligible via blocking analysis.

Lanczos results give $E_0^{(m_{\rm max})} = 0.429(20)$ for $m_{\rm max} = 47$ as shown in Fig.~\ref{fig:proton_correlator}; see the Supplemental Material for more details and pion results.
The two-sided residual-bound window is $[0.350(31), 0.523(26)]$.
A high-statistics benchmark is provided by model-averaged fits to correlation functions computed by the NPLQCD Collaboration using the same action with $N_{\rm cfg} = 670$  and 512 Gaussian-smeared sources per configuration:  $aM_N^{\rm big} = 0.4244(44)$.
Lanczos results agree within $1\sigma$.

\begin{figure}[t!]
                \centering
                \includegraphics[width=0.48\textwidth]{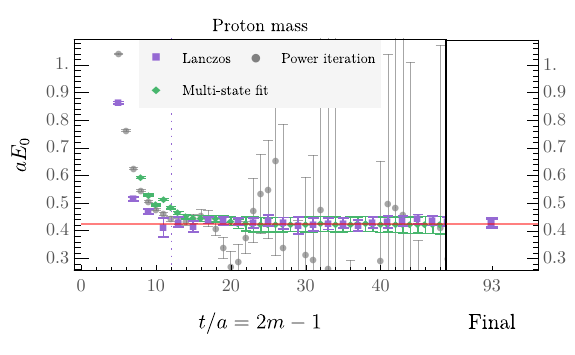}
                \includegraphics[width=0.48\textwidth]{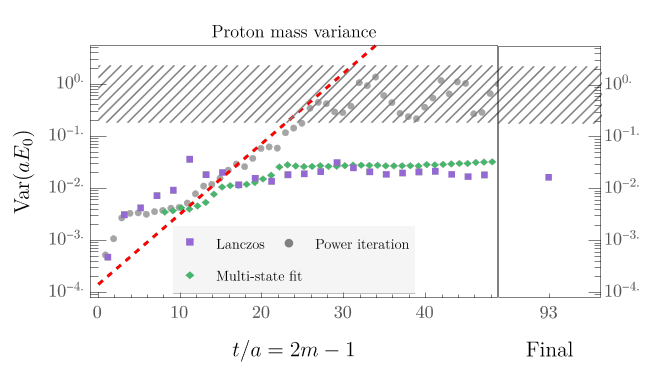}
            \caption{\label{fig:proton_correlator}
            Top: Proton $E(t)$, $\Em$, and multi-state fit results. $aM_N^{\rm big}$ is shown as a red band; other details are as in Fig.~\ref{fig:SHO_correlator}.
            Bottom: variances compared with Parisi-Lepage scaling $\propto e^{(M_N - \frac{3}{2} m_\pi)t}$~\cite{Parisi:1983ae,Lepage:1989hd} (red dashed line). Hatching indicates the noise region where $\text{SNR} \leq 1$ and variance estimates may be unreliable, which ranges from $\text{Var}E_0^{(m)} = [E_0^{(m)}]^2$ to the finite-statistics variance limit  $N_{\rm src} \text{Var}\, E(a)$~\cite{Wagman:2016bam}, where assuming negligible autocorrelations $N_{\rm src} = 64 \times 76$.}
\end{figure}

Two-state fits to $C_p(t)$ with $t_{\rm min}/a = 11$ give $0.4383(86)$ with $\chi^2/\text{dof} = 0.71$.
A weighted average of multi-state fits with all $t_{\rm min}/a \geq 2$ using the strategy in Refs.~\cite{NPLQCD:2020ozd,Amarasinghe:2021lqa} gives $0.443(12)$.
These are $1.5\sigma$ and $1.6\sigma$ larger than $aM_N^{\rm big}$, respectively.
For very large $t_{\rm max} \in [25,48]$, multi-state fits give accurate results $0.420(30)$ with larger uncertainties.
Lanczos energy estimators converge to exact results within $1\sigma$ for $2m-1 \gtrsim 11$, while multi-state fit weighted averages (with $t_{\rm plateau} \geq 5$~\cite{NPLQCD:2020ozd} to exclude short one-state fits) do not converge to $M_N^{\rm big}$ within $1\sigma$ until $t_{\rm max} \gtrsim 20$.
For $11 \lesssim t_{\rm max}/a \lesssim 20$, multi-state fits show $\geq 1\sigma$ signals of bias, effective-mass results have large uncertainties that appear non-Gaussian, while Lanczos results achieve negligible bias with comparable variance to multi-state fit results.

The variance of $E_0^{(m)}$ is shown in Fig.~\ref{fig:proton_correlator}. 
It approaches an $O(10^{-2})$ constant for $t/a \gtrsim 20$ with qualitatively similar behavior to the multi-state fit variance.
This is in contrast to power-iteration/effective-mass results, whose correlations do not saturate and variance grows exponentially with Parisi-Lepage scaling.
Lanczos therefore provides an algebraic estimator that is free from exponential SNR degradation.
Perhaps relatedly, correlations $\text{Corr}[E_0^{(m)}, E_0^{(m')}] \sim [0.5,1]$ are visible for $m,m' \gtrsim 20$ that suggest Lanczos results using bootstrap median estimators are relatively insensitive to $C(t)$ with $t/a \gtrsim 40$.
Significant large-$m$ correlations are specific to boostrap-median Lanczos estimators; sample-mean estimators have smaller correlations but larger variance as shown in the Supplemental Material.
In both cases, Lanczos energy estimators avoid the finite-sample bias seen in very large-$t$ power-iteration/effective-mass results, for which the variance is bounded to be $\geq 1/N$ times the variance of $E(a)$ due to circular statistics~\cite{Fisher:1993} of correlation-function phases and $\text{Re} E(t) \rightarrow \frac{3}{2}m_\pi$ in the $t \rightarrow \infty$ limit at finite $N$~\cite{Wagman:2016bam}.

{\it \bf Discussion:}
The method for applying the Lanczos algorithm to infinite-dimensional transfer matrices introduced here provides accurate predictions for simple harmonic oscillator and LQCD ground-state energies.
In particular, Lanczos achieves higher accuracy and similar precision to multi-state fits involving small imaginary times.
A two-sided error bound further shows excited-state effects cannot shift Lanczos results far outside their statistical uncertainties. 
Challenges arise from spurious eigenvalues, but an implementation of the Cullum-Willoughby test using bootstrap histograms and bootstrap median estimators mitigate their effects in these examples.
Future refinements improving the robustness of spurious eigenvalue removal could improve the statistical precision of Lanczos energy results and residual bounds.

The fact that Lanczos SNR approaches a constant at large $m$ can be understood from the perspective of the Lanczos algorithm as a Krylov-space projection method~\cite{Parlett}.
Explicitly, projection operators can be constructed from the Ritz vectors as
    $P_n^{(m)} \equiv \big| y_n^{(m)} \bigr> \bigl< y_n^{(m)} \bigr|$ and are related to Ritz values as
\begin{equation}
\begin{split}
  \lambda_n^{(m)} &= \mbraket{y_n^{(m)}}{T}{y_n^{(m)}} =  \frac{\mbraket{\psi}{P_n^{(m)} T P_n^{(m)}}{\psi}}{|\omega_{1n}^{(m)}|^2 \braket{\psi}{\psi}},
\end{split}
\end{equation}
for a Hermitian subspace of non-spurious Ritz values~\cite{Hackett:2024xnx}.
$P_n^{(m)}$ converges to the projection operator $P_n \equiv \ket{n}\bra{n}$ with finite-$m$ corrections analogous to those appearing the KPS eigenvalue bound, e.g. $P_n^{(m)} = P_n + O(e^{-4 m \sqrt{\delta}})$ for $\delta \ll 1$
~\cite{Parlett,Saad:1980,Saad:1982}. 
This provides the Hilbert-space operator relation
$\psi P_n T P_n \psi^\dagger = |Z_n|^2 e^{-a E_n} \left| n \right> \left< n \right| + O(e^{-4 m \sqrt{\delta}})$.
At finite statistics, $P_n^{(m)}$ depends on the gauge-field ensemble used and $\psi P_n^{(m)} T P_n^{(m)} \psi^\dagger$ provides an observable whose finite-statistics expectation value coincides with $\lambda_n^{(m)}$.
This suggests the variance of $\lambda_n^{(m)}$ is proportional to that of $\psi P_n^{(m)} T P_n^{(m)} \psi^\dagger$ in the limit of large statistics.
If the statistical ensemble and $m$ are both large, then $P_n^{(m)}$ approaches $P_n$ and in this limit the variance of $\lambda_n^{(m)}$ is 
proportional to the variance of the projector $\left|n \right> \left< n \right|$.

The variances of projectors have appealing features including the non-appearance of states with different quantum numbers than the signal squared (e.g. three-pion states in nucleon variances)~\cite{DellaMorte:2007zz,DellaMorte:2008jd,DellaMorte:2010yp}; the finiteness of the variance of $\left|n \right> \left< n \right|$ is already sufficient to ensure the variance of $\lambda_n^{(m)}$ approaches a finite $m$-independent value.
Analogous arguments slow the variance of $\omega_{1n}^{(m)}$ is proportional to that of $\psi P_n^{(m)} \psi^\dagger$ and likewise approaches a constant at large $m$.
The SNR of $E_n^{(m)} = -1/a \ln \lambda_n^{(m)}$ therefore approaches a non-zero constant for large $m$.
This should be contrasted with the power-iteration method, for which $E(2ma)$ is defined by a log-ratio of $C(2ma)$ and $C((2m-1)a)$, both of which have zero SNR at large $m$.
Since they are not perfectly correlated, the SNR of $E(2ma)$ vanishes (exponentially) at large $m$.

In conclusion, Lanczos provides rapidly convergent algebraic energy estimators without SNR problems and two-sided bounds
on excited-state effects that 
could be useful for a wide range of hadron spectroscopy calculations where isolating ground states is challenging, including studies of nucleon, nuclear, and highly boosted systems.
The existence of two-sided bounds on Lanczos energy estimates using a finite number of $C(t)$ points provides a qualitative improvement over the one-sided bounds available to variational methods and could have wide-ranging applications.

\begin{acknowledgments}
  I thank Dan Hackett, Ryan Abbot, Beno{\^i}t Assi, Zohreh Davoudi, Will Detmold, George Fleming, Dorota~Grabowska, Anthony Grebe, Marc Illa, Will~Jay, Andreas Kronfeld, Assumpta Parre{\~n}o, Robert Perry, Dimitra Pefkou, Fernando Romero-L{\'o}pez, ~~~~~~~~~~~~~~~~~~~~~Martin~Savage, Phiala Shanahan, and Ruth Van de Water for stimulating discussions and helpful comments.
This manuscript has been authored by Fermi Research Alliance, LLC under Contract No.\ DE-AC02-07CH11359 with the U.S.\ Department of Energy, Office of Science, Office of High Energy Physics.
This research used facilities of the USQCD Collaboration, which are funded by the Office of Science of the U.S. Department of Energy.
The Chroma~\cite{Edwards:2004sx},  QUDA~\cite{Clark:2009wm,Babich:2011np,Clark:2016rdz}, and QDP-JIT~\cite{6877336} software libraries were used in this work. Numerical analysis was performed and figures were produced using Mathematica \cite{Mathematica}.

\end{acknowledgments}

\bibliography{Lanczos_bib}

\clearpage

\section*{Supplementary material}

This Supplementary Material provides additional details on the implementation of the Lanczos algorithm used to compute the results of the main text. A proof that thermal effects can be described without modifying the basic algorithm is presented in Sec.~\ref{sec:thermal}. An implementation of the Cullum-Willoughby method for removing spurious eigenvalues is introduced and its numerical application described in Sec.~\ref{sec:spurious}. An oblique Lanczos method suitable for describing noisy Monte Carlo results leading to non-Hermitian $T^{(m)}_{ij}$ is presented in Sec.~\ref{sec:oblique}. Finally, Sec.~\ref{sec:plots} presents additional numerical results on the residuals and correlations of Lanczos results.

\subsection{Thermal effects}\label{sec:thermal}

Consider a function $f(t)$ sampled at discrete points $t \in \{a,2a,\ldots,\beta a\}$ with a spectral representation
\begin{equation}\label{eq:thermal_spec}
    f(t) = \sum_{n=0}^\infty X_n e^{-E_n t} + Y_n e^{F_n t}, 
\end{equation}
where $E_n, F_n > 0$ and $X_n,Y_n \in \mathbb{R}$.
Particular cases include bosonic thermal correlation functions in which $Y_n = X_n e^{-\beta E_n}$ and $F_n = E_n$, as well as fermionic thermal correlation functions in which the $F_n$ are equal to energies of fermionic states with opposite parity to those with energies $E_n$. 
The thermal states with transfer matrix eigenvalues $e^{a F_n}$ can be treated as additional states on equal footing with those with eigenvalues $e^{-a E_n}$ by defining
\begin{equation}
    f(t) = \sum_{k=0}^\infty Z_k e^{L_k t},
\end{equation}
where
\begin{equation}
    \begin{split}
        Z_k \equiv \begin{cases}
            X_{k/2}, & k \text{ even} \\
            Y_{(k-1)/2}, & k \text{ odd}
        \end{cases}, \\
        L_k \equiv \begin{cases}
            -E_{k/2}, & k \text{ even} \\
            F_{(k-1)/2}, & k \text{ odd}
        \end{cases}.
    \end{split}
\end{equation}
Using this representation as a starting point, a matrix $T$ can be defined that acts analogously to the physical transfer matrix in the zero-temperature case.

A set of vectors $\ket{k}$ and dual vectors $\bra{k}$ can be defined to have inner product $\braket{k}{k'} = \delta_{kk'}$.
A vector $\ket{\psi}$ can be defined by
\begin{equation}
    \braket{k}{\psi} \equiv \sqrt{Z_k},
\end{equation}
where the branch cut of $\sqrt{Z_k}$ is placed along the negative imaginary axis in case $Z_k$ is negative.
The dual vector $\bra{\chi}$ can be defined by
\begin{equation}
    \braket{\chi}{k} \equiv \sqrt{Z_k},
\end{equation}
An operator $T$ can be defined by\begin{equation}
    \mbraket{k}{T}{k'} = e^{a L_k} \delta_{kk'}.
\end{equation}
Since $L_k \in \mathbb{R}$, $T$ is represented by a real and diagonal matrix in the $\ket{k}$ basis and is therefore a Hermitian operator.
This provides a matrix-element representation for $f(t)$ as
\begin{equation}
\begin{split}
    \mbraket{\chi}{T^{t/a}}{\psi} &= \sum_{k,k'} \braket{\chi}{k}\mbraket{k}{T^{t/a}}{k'} \braket{k'}{\psi} \\
    &= \sum_{k,k'} \sqrt{Z_k} e^{L_k t} \delta_{kk'} \sqrt{Z_{k'}} \\
    &= \sum_{k} Z_k e^{L_k t} \\
    &= f(t).
\end{split} 
\end{equation}

Regardless of whether $f(t)$ can be associated with a physical correlation function, this provides a representation of $f(a)$, $f(2a)$, ..., $f(t)$ as matrix elements $\mbraket{\chi}{T}{\psi}$, $\mbraket{\chi}{T^2}{\psi}$, ..., $\mbraket{\chi}{T^{t/a}}{\psi}$ of a Hermitian operator $T$.
Application of the Lanczos algorithm described here to $f(t)$ will lead to iterative approximations $\lambda_k^{(m)}$ that converge to eigenvalues $e^{aL_k}$ of $T$ in the limit of large Lanczos steps $m \rightarrow \infty$.
This implies
\begin{equation}
\begin{split}
    \{ \lambda_k^{(m)}  \} \rightarrow& \{ e^{aL_k} \} = \{ e^{-a E_n},\ e^{a F_n}  \},
    \end{split}
\end{equation}
and the estimators $E_k^{(m)} = -(1/a) \ln \lambda_k^{(m)}$ therefore converge to
\begin{equation}
\begin{split}
    \{ E_k^{(m)}  \} &\rightarrow \{ E_n,\ -F_n  \}.
    \end{split}
\end{equation}
The ``energy spectrum'' obtained by applying Lanczos methods to $f(t)$ for $m$ steps will therefore approximate some admixture of the $E_n$ and $-F_n$.

Applying Lanczos methods to bosonic thermal correlation functions  will therefore result in positive $E_n^{(m)}$ approximating physical energies $E_n$ and negative $E_n^{(m)}$ approximating $-E_n$.
Applying Lanczos methods to fermionic thermal correlation functions will result in  positive $E_n^{(m)}$ approximating the energies of fermionic states with the same parity as $\ket{\psi}$ and negative $E_n^{(m)}$ approximating minus the energies of states with opposite parity.

Since this construction applies even when some of the $X_n$ are negative, it is valid even when $f(t)$ describes an ``asymmetric'' correlation function represented as $\left< \mathcal{O}'(t) \mathcal{O}^\dagger(0)\right>$ with $\mathcal{O} \neq \mathcal{O}'$ in the physical Hilbert space.
An oblique Lanczos method suitable for asymmetric correlation functions is described in Sec.~\ref{sec:oblique} below.

\begin{figure*}[t!]
                \centering
                \includegraphics[width=0.48\textwidth]{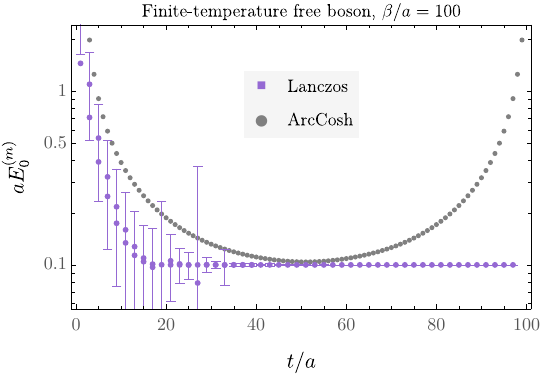}
                \includegraphics[width=0.48\textwidth]{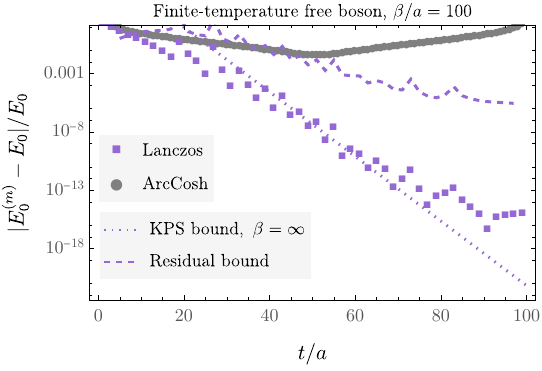}
            \caption{\label{fig:mock_correlator_boson}
            Comparison of the  estimator $\text{arccosh}([C(t+2a)+C(t)]/[2C(t+a)])$ with Lanczos-based estimators for a time series with $\beta/a = 100$ points corresponding to Eq.~\eqref{eq:thermal_spec} with $X_n = Y_n = (n+1)$ and $a E_n = 0.1 (n+1)$ for $n=0,\ldots,50$. As in Fig.~\ref{fig:mock_correlator}, estimators $E_0^{(k)}$ after $k$ iterations of Lanczos are shown at $t/a = 2k-1$ since this is the largest $t/a$ correlation function involved in its calculation. The residual bound computed via the Lanczos algorithm with Eq.~\eqref{eq:Rbound} is shown as a dashed line. The ground-state version of the KPS bound, Eq.~\eqref{eq:KPS0}, only applies for $\beta \rightarrow \infty$ limit and can be violated when thermal effects in $f(t)$ are significant.
                }
\end{figure*}

The practical convergence of Lanczos-based estimators for time series of the form Eq.~\eqref{eq:thermal_spec} that can be interpreted as thermal correlation functions with $\beta/a = 100$ are shown in Figs.~\ref{fig:mock_correlator_boson}. 
Thermal states show up as negative-energy eigenvalues in Lanczos results for $t/a \gtrsim 20$.
In the same region, the approach of the Lanczos estimator to the ground-state energy is non-monotonic.
The residual bound applies for all $t/a$ and provides a rigorous systematic uncertainty on the distance between $E^{(m)}_k$ and some energy eigenvalue.
The  ground-state version of the KPS bound in the form of Eq.~\eqref{eq:KPS0} only applies for $\beta \rightarrow \infty$ --- empirically it holds for $t \lesssim 3/4 \beta$ in this example but can be seen to be violated for larger $t$.

A bound valid at finite $\beta$ can be obtained from the general form of the KPS bound~\cite{Kaniel:1966,Paige:1971,Saad:1980},
\begin{equation}\label{eq:KPS}
    0 \leq \frac{ \lambda_n  - \lambda_n^{(m)} }{ \lambda_n - \lambda_{\infty} } \leq \left[ \frac{ K_n^{(m)} \tan \phi_n}{ T_{m-n-1}(\Gamma_n)} \right]^2,
\end{equation}
where the smallest eigenvalue of the transfer matrix is denoted $\lambda_\infty$ and converges to zero for a lattice gauge theory with an infinite-dimensional Hilbert space such as LQCD,  $\cos\phi_n \equiv \braket{n}{\psi} = Z_n$,
\begin{equation}
\begin{split}
    \Gamma_n &\equiv 1 + \frac{2(\lambda_n - \lambda_{n+1})}{\lambda_{n+1} - \lambda_{\infty}} = 2 e^{a(E_{n+1}-E_n)} - 1,
    \end{split}
\end{equation}
and
\begin{equation}
    K_n^{(m)} \equiv \prod_{l=1}^{n-1} \frac{\lambda_l^{(m)} - \lambda_{\infty}}{\lambda_l^{(m)} - \lambda_l}, \hspace{20pt} n > 0,
\end{equation}
with $K_0^{(m)} \equiv 1$.
However, this bound requires knowledge of the entire spectrum to compute.
Only for the case of $n=0$ does the bound reduce to an expression in terms of only $E_n$ and $Z_n$, which results in Eq.~\eqref{eq:KPS0}.

\subsection{Spurious eigenvalues}\label{sec:spurious}

Implementing the Cullum-Willoughby procedure~\cite{Cullum:1981,Cullum:1985} for removing spurious eigenvalues in Lanczos applications to noisy Monte Carlo results requires the definition of a threshold for when eigenvalues of $\Tm_{ij}$ and $\CWm_{ij}$ should be considered identical, where $\CWm_{ij}$ is the matrix obtained by removing the first row and column from the tridiagonal matrix $\Tm_{ij}$.
The threshold should only be different from zero because of statistical uncertainties (and to a lesser extent finite-precision arithmetic used when generating field configurations and/or executing the Lanczos algorithm).
An automated procedure for choosing this threshold based on bootstrap eigenvalue histograms is defined below.
It introduces two hyperparameters: $\Delta$ controls the number of histogram bins and $K_{\rm CW}$ is an $O(1)$ tolerance parameter specifying how many samples are required to call a bin ``non-spurious'' as described below.

Since $T$ has only positive eigenvalues by assumption, non-positive eigenvalues of $\Tm_{ij}$ must be spurious eigenvalues arising from statistical noise.
Using the oblique Lanczos formalism described in Sec.~\ref{sec:oblique} below, the $\Tm_{ij}$ are all real and the Ritz values $\lambda_k^{(m)}$ are therefore either exactly real or have non-zero imaginary parts and come in complex conjugate pairs.\footnote{Using the symmetric Lanczos algorithm with complex $\beta_j$ introduces small but non-zero imaginary parts to all eigenvalues and a threshold for distinguishing positive from non-positive eigenvalues must be introduced.}
All Ritz values that have non-zero imaginary parts at working precision can therefore be discarded as spurious (in the numerical examples here I discard Ritz values with $|\text{arg}(\lambda_k^{(m)})| > 10^{-12}$).
Using arbitrary-precision arithmetic instead of double precision leads to significant changes to complex Ritz values but provides consistent results for positive Ritz values, and in the Monte Carlo analyses here I have adopted double precision for convenience.\footnote{It is likely that calculations employing significantly larger imaginary time extents and Lanczos iteration counts than those studied here would need to employ higher precision during execution of the Lanczos algorithm.}

Bootstrap resampling~\cite{Efron:1982,Davison:1997,Young:2014} can help to identify spurious eigenvalues because they arise due to statistical noise and their values are therefore more sensitive to noise than non-spurious eigenvalues.
Ritz values are computed for sample mean correlation functions and for each of $N_{\text{boot}}$ ensembles obtained using bootstrap resampling~\cite{Efron:1982,Davison:1997,Young:2014}. 
In the numerical results of this work, $N_{\rm boot} = 200$ is used throughout.

The bootstrap Ritz value distribution can be analyzed using a variation of the Cullum-Willoughby criterion:
\begin{itemize}
  \item Compute the Ritz values $\lambda_k^{(b,m)}$ with $k \in \{1,\ldots,m\}$ an arbitrary ordering, $m \in \{1,\ldots,N_{\rm it}\}$ where $N_{\rm it} = (\beta+1)/2$ is the maximum number of Lanczos iterations, and $b \in \{1,\ldots,N_{\rm boot}\}$.
  \item Discard any $\lambda_k^{(b,m)}$ with non-zero imaginary parts as spurious.
  \item Compute the eigenvalues $\tau_k^{(b,m)}$ of the matrices $\CWbm_{ij}$ obtained by removing the first row and column of $\Tbm_{ij}$ for each $m$ and $b$.
  \item Compute $d_k^{(b,m)} \equiv \text{min}_{j\in\{1,\ldots,m-1\}} | \lambda_k^{(b,m)} - \tau_j^{(b,m)}|$ for all $k$, $m$, and $b$.
    \item Denote the total number of approximately positive eigenvalues computed across all bootstrap ensembles by $N_+$. Define the number of eigenvalues that could be determined if these were all non-spurious by $N_\lambda \equiv \text{round}[N_+ / N_{\rm boot} / N_{\rm it}]$ where round denotes rounding to the nearest integer.
    \item Histogram $\ln d_k^{(b,m)}$ with a number of bins defined by $N_{\rm bins} \equiv \Delta N_{\lambda}$ where $\Delta$ is another hyperparameter. The energy results and uncertainties studied here are insensitive to $\Delta$ over the range $\Delta \sim 2 - 10$. For concreteness I take $\Delta = 4$ everywhere.
    \item Define $\delta_{CW} \equiv N_{\rm boot} (N_{\rm it} - N_{\lambda}) K_{\rm CW} / \Delta$, the number of counts required to indicate a bin contains an eigenvalue that is repeated across Lanczos iterations and bootstrap ensembles and is therefore not spurious. Here, $K_{\rm CW}$ is a $O(1)$ hyperparameter discussed more below; I take $K_{\rm CW} = 3$ everywhere.
    \item Find the first histogram bin (ordered with $\ln d_k^{(b,m)}$  increasing) with more than $\delta_{CW}$ counts. Denoting this histogram bin by  $\ln d_k^{(b,m)} \in [B_i, B_{i+1}]$, define the Cullum-Willoughby threshold by $\varepsilon_{\rm CW} \equiv e^{B_{i}} / F_{CW}$ where $F_{CW}$ is the last hyperparameter. The energy results and uncertainties studied here are insensitive to $F_{CW}$ over the range $F_{CW} \sim 5 - 50$. For concreteness I take $F_{CW} = 50$ everywhere.
\end{itemize}
The result of this bootstrap histogram analysis is the Cullum-Willoughby threshold $\varepsilon_{\rm CW}$.
This threshold is used to remove spurious Ritz values:
\begin{itemize}
    \item Compute the sample mean Ritz values, $\lambda_k^{(m)}$, and discard $\lambda_k^{(m)}$ with non-zero imaginary parts. Compute the eigenvalues $\tau_k^{(m)}$ and distances $d_k^{(m)}$. Remove $\lambda_k^{(m)}$ with $d_k^{(m)} > \varepsilon_{\rm CW}$ as spurious.
    \item Remove Ritz values with $\lambda_k^{(m)} > 1$, which correspond to thermal states as described in Sec.~\ref{sec:thermal}.
   \item The remaining  non-spurious Ritz values $\lambda_k^{(m)} < 1$ are sorted as $\lambda_0^{(m)} > \lambda_1^{(m)} > \ldots$ for each $m$.
\end{itemize}
These provide Lanczos energies $E_k^{(m)} \equiv -(1/a)\ln \lambda_k^{(m)}$ and in particular the central values for $E_0^{(m)}$ used here.

\begin{figure*}[t!]
                \centering
                \includegraphics[width=0.48\textwidth]{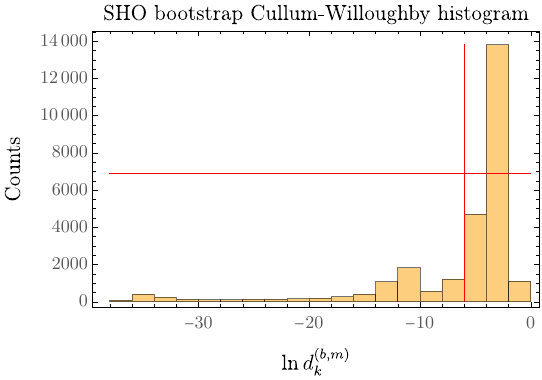}
                \includegraphics[width=0.48\textwidth]{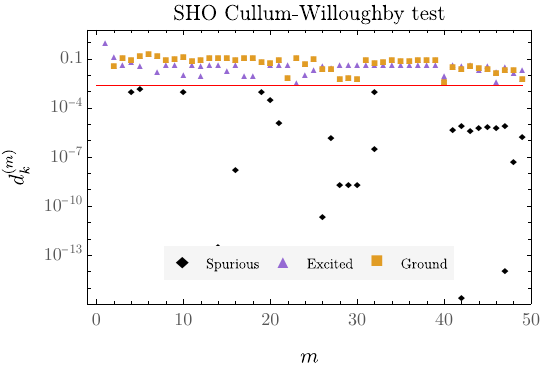}
                \includegraphics[width=0.48\textwidth]{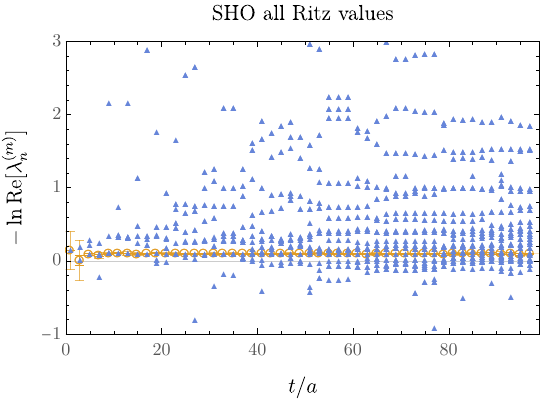} 
                \includegraphics[width=0.48\textwidth]{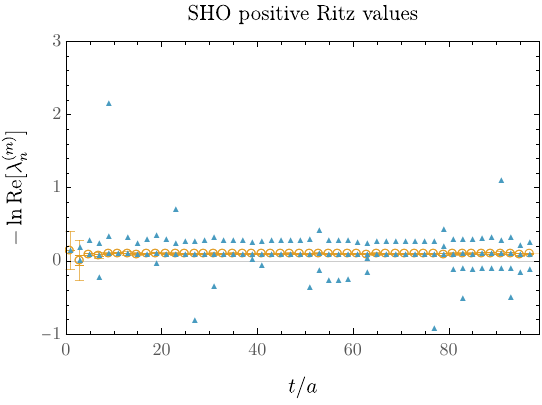} 
                \includegraphics[width=0.48\textwidth]{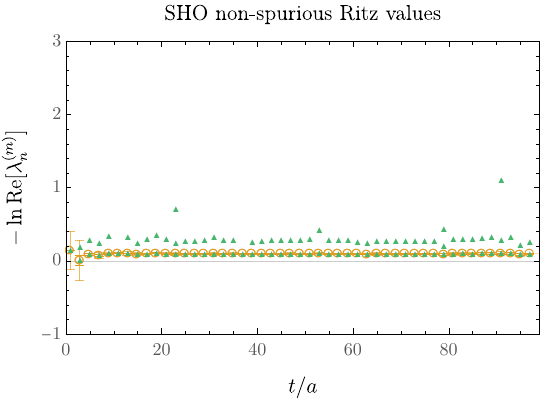} 
            \caption{\label{fig:SHO_spurious}
            Summary of the Cullum-Willoughby test used to identify and remove spurious eigenvalues for complex scalar field theory results. The bootstrap histogram of $\ln d_k^{(b,m)}$ is shown top-left with $N_{\lambda}$ computed using $\Delta = 4$ and $K_{\rm CW} = 3$ shown as a horizontal red line. The vertical red line shows the corresponding value of $\ln \varepsilon_{\rm CW}$ computed as described in the text. Top-right, sample mean results for $d_k^{(m)}$ are shown in comparison with $\varepsilon_{\rm CW}$ (horizontal red line) with those corresponding to spurious, non-spurious excited-state, and non-spurious ground-state eigenvalues shown as black diamonds, purple triangles, and orange squares, respectively. Middle-left, all of the spurious and non-spurious eigenvalues are shown along with the results for $E_0^{(m)}$ shown in the main text. Analogous comparisons are shown for all the (spurious and non-spurious) positive eigenvalues, middle-right, and all of the non-spurious eigenvalues (excluding those associated with thermal states), bottom.
                }
\end{figure*}

\begin{figure*}[t!]
                \centering
                \includegraphics[width=0.48\textwidth]{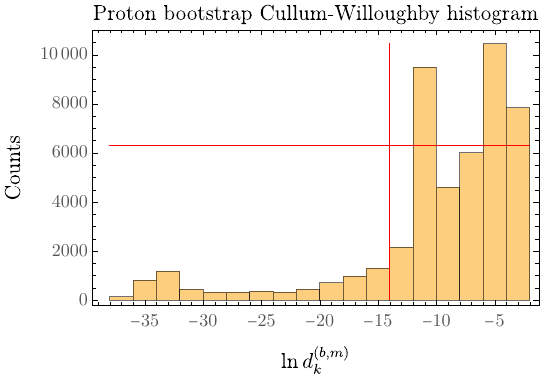}
                \includegraphics[width=0.48\textwidth]{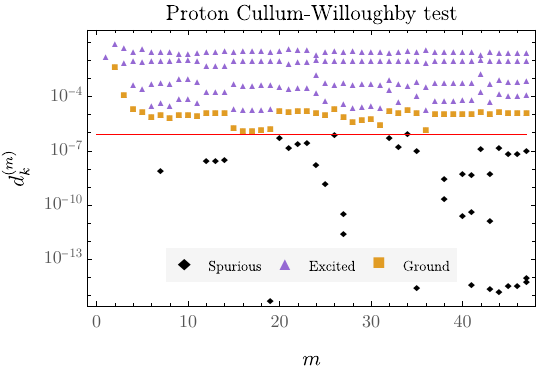}
                \includegraphics[width=0.48\textwidth]{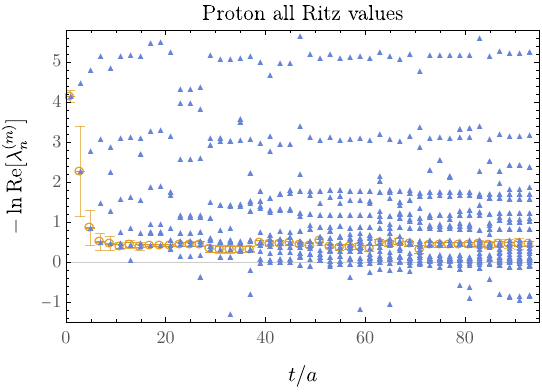} 
                \includegraphics[width=0.48\textwidth]{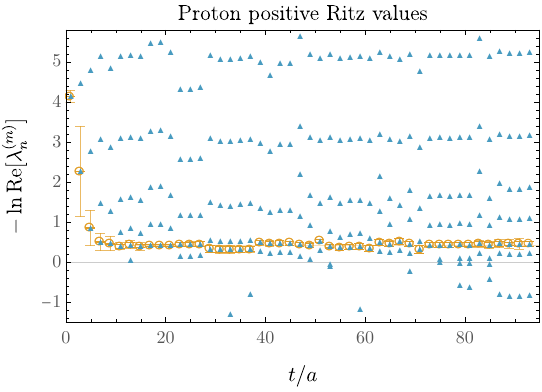} 
                \includegraphics[width=0.48\textwidth]{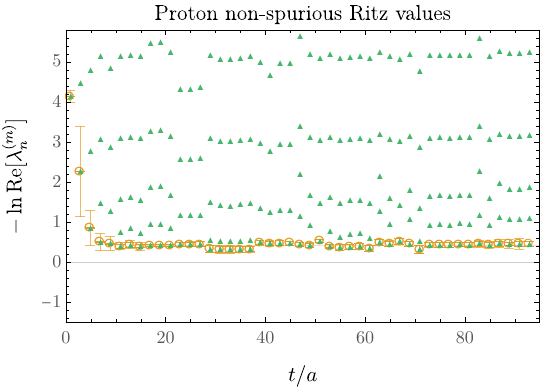} 
            \caption{\label{fig:proton_spurious}
            Summary of the Cullum-Willoughby test for the LQCD proton results in the main text. Details are as in Fig.~\ref{fig:SHO_spurious}.
                }
\end{figure*}

\begin{figure*}[t!]
                \centering
                \includegraphics[width=0.48\textwidth]{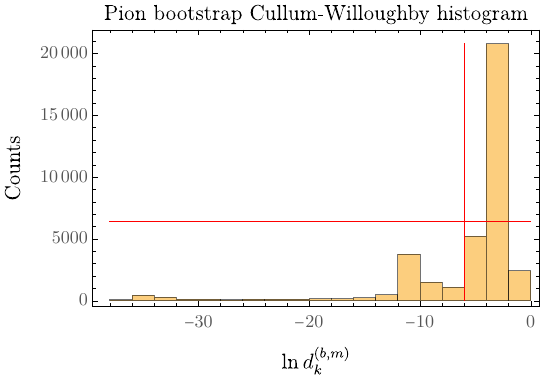}
                \includegraphics[width=0.48\textwidth]{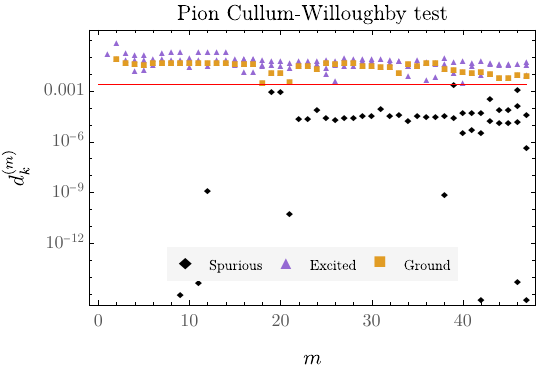}
                \includegraphics[width=0.48\textwidth]{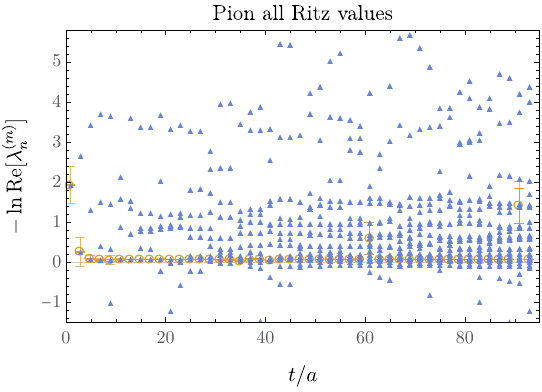} 
                \includegraphics[width=0.48\textwidth]{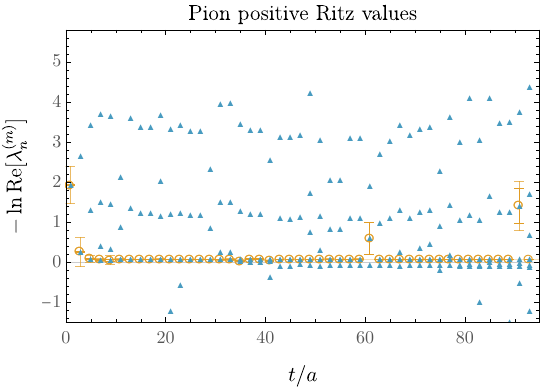} 
                \includegraphics[width=0.48\textwidth]{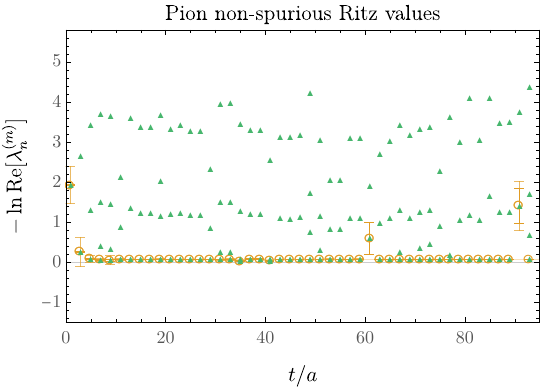} 
            \caption{\label{fig:pion_spurious}
            Summary of the Cullum-Willoughby test for the LQCD pion results in Sec.~\ref{sec:plots}. Details are as in Fig.~\ref{fig:SHO_spurious}.
                }
\end{figure*}

Example results for the determination of Cullum-Willoughby thresholds and spurious eigenvalue identification for SHO, LQCD proton, and LQCD pion results are shown in Figs.~\ref{fig:SHO_spurious}-\ref{fig:pion_spurious}.
It is noteworthy that the gap in $d_n^{(m)}$ between spurious and non-spurious eigenvalues is not large, and in some cases the non-spurious eigenvalue associated with the ground state is close to the threshold.
This means that a small increase in $K_{\rm CW}$ can cause the ground-state eigenvalue for a particular Lanczos iteration to be labeled spurious, at which point the largest non-spurious eigenvalue will correspond to an excited-state energy.
Conversely, a small decrease in $K_{\rm CW}$ can cause a spurious eigenvalue with larger magnitude than $\lambda_0^{(m)}$ to be labeled non-spurious, at which point an unphysically small value for $E_0^{(m)}$ will be obtained.
Although this can introduce undesirable numerical instabilities to the determinations of $E_0^{(m)}$ for particular $m$, spurious eigenvalues --- by their spurious nature --- have broadly distributed $\ln d_n^{(m)}$ that only rarely approach the $\ln d_n^{(m)}$ of non-spurious eigenvalues.
When spurious eigenvalues do have $\ln d_n^{(m)}$ close to those non-spurious eigenvalues, they are often (although not always) also close in magnitude, and in these cases similar $E_0^{(m)}$ are obtained regardless of which eigenvalues are labeled spurious.
The appearance of such ``multiple eigenvalues'' is commonplace in applications of Lanczos to finite matrices with floating-point arithmetic~\cite{Cullum:1981,Cullum:1985} and complicates Lanczos determinations of excited-state energies.

To understand whether a small change in $K_{\rm CW}$ could lead to a significant change in $\varepsilon_{\rm CW}$ and therefore to a change in whether eigenvalues with relatively large magnitudes are labeled as spurious, it is sufficient to examine the bootstrap histogram of $\ln d_n^{(m)}$.
If $\varepsilon_{\rm CW}$ is near the top of a histogram bin that has smaller $\ln d_n^{(m)}$ than all other bins with greater than $\varepsilon_{\rm CW}$ counts, then a small change in $K_{\rm CW}$ can cause $\varepsilon_{\rm CW}$ to be associated with either this bin or a different bin.
In this case, it is worthwhile to study the variation of $E_0^{(m)}$ results with $K_{\rm CW}$ choices that lead to both possibilities.
Any significant effects on fit results $E_0$ and/or $\delta E_0$ arising from such variations should be considered additional systematic uncertainties.
In the examples studied here, varying $\Delta$, $K_{CW}$, and $F_{CW}$ over the ranges indicated leads to negligible effects, and I do not associate a systematic uncertainty with this variation.

\subsection{Oblique Lanczos}\label{sec:oblique}

A nonsymmetric, or oblique, version of the Lanczos algorithm can be used to compute the eigenvalues of non-Hermitian matrices~\cite{Saad:1982,Parlett:1985,Nachtigal:1993}.
Even though the transfer matrix $T$ can be assumed to be Hermitian for LQCD applications, the oblique Lanczos formalism is needed to describe fermions at non-zero temperature as well as asymmetric correlation functions
\begin{equation}
    C_{\chi \psi}(t) \equiv \left< \chi(t) \psi^\dagger(0) \right> = \mbraket{\chi}{T^{t/a}}{\psi} + \ldots.
\end{equation}
The oblique Lanczos formalism also avoids the complex $\Tm_{ij}$ that arise in applications of the symmetric Lanczos algorithm to noisy Monte Carlo results and violate the usual assumptions of theoretical Lanczos analyses.
Oblique Lanczos therefore provides a theoretically rigorous, as well as more numerically stable, starting point for applying Lanczos in situations when $T$ is Hermitian but the Hermiticity of $\Tm$ is broken at finite statistics. 
All numerical results in this work use the oblique Lanczos algorithm in practice.

The primary difference between oblique and symmetric Lanczos is that the former has distinct Krylov spaces of left- and right-Lanczos vectors.
These vectors will be denoted $\bra{w_n}$ and $\ket{v_n}$ respectively.
The first iteration of oblique Lanczos involves the matrix element
\begin{equation}
    \begin{split}
        \alpha_1 \equiv \mbraket{w_1}{T}{v_1} = \frac{\mbraket{\chi}{T}{\psi}}{\braket{\chi}{\psi}} = \hat{C}_{\chi\psi}(1),
    \end{split}
\end{equation}
where the last equality shows the connection to the ratio of correlation functions $\hat{C}_{\chi\psi}(t) \equiv C_{\chi\psi}(t) / C_{\chi\psi}(0)$ and implies that $m=1$ results coincide with the  effective mass (and symmetric Lanczos).
The base cases for other tridiagonal matrix elements defined below are $\beta_1 = 0$ and $\gamma_1 = 0$.
The matrix elements needed for a recursive construction of oblique Lanczos that does not explicitly access the Lanczos vectors are
\begin{equation}\label{eq:lr_MEs}
    \begin{split}
        A_j^{(k)} &\equiv \mbraket{ w_j }{ T^k }{ v_j }, \\
        G_j^{(k)} &\equiv \mbraket{ w_j }{ T^k }{ v_{j-1} }, \\
        B_j^{(k)} &\equiv \mbraket{ w_{j-1} }{ T^k }{ v_j }.
    \end{split}
\end{equation}

\subsubsection{Recursion relations}

After the first Lanczos iteration, the residuals for the left- and right-Lanczos vectors are defined by~\cite{Saad:1982,Parlett:1985,Nachtigal:1993}
\begin{equation}\label{eq:lr_vecs}
\begin{split}
    \ket{r_{j+1}} \equiv (T - \alpha_j) \ket{v_j}  -  \beta_j \ket{v_{j-1}}, \\
    \bra{s_{j+1}} \equiv \bra{w_j} (T - \alpha_j) - \gamma_j \bra{w_{j-1}}.
\end{split}
\end{equation}
The left- and right-Lanczos vectors for the next iteration are normalized versions of the corresponding residuals
\begin{equation}\label{eq:lr_norms}
\begin{split}
  \ket{v_{j+1}} &\equiv \frac{1}{\rho_{j+1} } \ket{r_{j+1}}, \\
    \bra{w_{j+1}} &\equiv \frac{1}{\tau_{j+1} } \bra{s_{j+1}}.
\end{split}
\end{equation}
Several choices for the normalization factors $\rho_j$ and $\tau_j$ are possible~\cite{Saad:1982,Parlett:1985,Nachtigal:1993}; in order
to simply subsequent expressions for tridiagonal matrix elements it is convenient to choose~\cite{Saad:1982,Parlett:1985}
\begin{equation}
    \rho_j \equiv \sqrt{|\braket{s_j}{r_j}|}, \hspace{20pt} \tau_j \equiv \frac{\braket{s_j}{r_j}}{\rho_j}. \label{eq:rhotau}
\end{equation}
In conjunction with the bi-orthogonality of the Lanczos vectors discussed further below, this choice leads to
\begin{equation}
    \braket{w_i}{v_j} = \delta_{ij}.
\end{equation}
The recursion relation required to compute $\braket{s_{j+1}}{r_{j+1}}$ in terms of matrix elements from previous iterations is given by
\begin{equation}\label{eq:sr}
    \begin{split}
      \braket{s_{j+1}}{r_{j+1}}  &= A_j^{(2)}  - 2 \alpha_j  A_j^{(1)} + \alpha_j^2  A_j^{(0)}  \\
        &\hspace{20pt}   + \alpha_j  \left(  \beta_j G_j^{(0)} + \gamma_j B_j^{(0)}  \right) \\
        &\hspace{20pt}  - \left( \beta_j G_j^{(1)} + \gamma_j  B_j^{(1)}  \right) \\
        &\hspace{20pt}  + \gamma_j\beta_j A_{j-1}^{(0)} ,
    \end{split}
\end{equation}
from which  $\rho_{j+1}$ and $\tau_{j+1}$ can be computed using Eq.~\eqref{eq:rhotau}.

Recursion relations for the matrix elements $A_j^{(k)}$, $B_j^{(k)}$, and $G_j^{(k)}$ can be derived as in the symmetric case by inserting Eqs.~\eqref{eq:lr_vecs}-\eqref{eq:lr_norms} into Eq.~\eqref{eq:lr_MEs}.
The recursion relation for $G_j^{(k)}$ is
\begin{equation}
    \begin{split}
        \tau_{j+1}  G_{j+1}^{(k)} &= 
            A_j^{(k+1)} - \alpha_j A_j^{(k)} - \gamma_j B_j^{(k)}.
            \end{split}
\end{equation}
The analogous recursion relation for $B_j^{(k)}$ is
\begin{equation}
    \begin{split}
        \rho_{j+1}  B_{j+1}^{(k)} &= 
            A_j^{(k+1)} - \alpha_j A_j^{(k)}  - \beta_j G_j^{(k)}.
    \end{split}
\end{equation}
The recursion relations for $A_j^{(k)}$ is
\begin{equation}\label{eq:diag_regular}
    \begin{split}
         \rho_{j+1} \tau_{j+1} A_{j+1}^{(k)}  &= A_j^{(k+2)}  - 2 \alpha_j  A_j^{(k+1)} + \alpha_j^2  A_j^{(k)}  \\
        &\hspace{20pt}   + \alpha_j  \left(  \beta_j G_j^{(k)} + \gamma_j B_j^{(k)}  \right) \\
        &\hspace{20pt}  - \left( \beta_j  G_j^{(k+1)} + \gamma_j  B_j^{(k+1)}  \right) \\
        &\hspace{20pt}  + \gamma_j\beta_j A_{j-1}^{(k)} .
    \end{split}
\end{equation}

After $m$ iterations of oblique Lanczos, the tridiagonal matrix
\begin{equation}
    \Tm_{ij} \equiv \mbraket{w_i}{T}{v_j}, 
\end{equation}
expressing matrix elements of $T$ in the Lanczos-vector basis is given by
\begin{equation} \label{eq:untri}
  \Tm_{ij} = \begin{pmatrix} \alpha_1 & \beta_2 &  & & & 0 \\ 
    \gamma_2 & \alpha_2 & \beta_3 & & & \\
    & \gamma_3 & \alpha_3 & \ddots & &  \\
    & & \ddots & \ddots & \beta_{m-1} & \\
    & & & \gamma_{m-1} & \alpha_{m-1} & \beta_m \\
  0 & & & & \gamma_m & \alpha_m \end{pmatrix}_{ij},
\end{equation}
where the elements $\alpha_j$, $\beta_j$, and $\gamma_j$ can be derived by combining the recursion relations in Eq.~\eqref{eq:lr_vecs} with the orthogonality condition $\braket{w_j}{v_i} = 0$ for $i\neq j$.
Combining $\braket{w_{j+1}}{v_j} = \braket{w_j}{v_{j+1}} = 0$ with the recursion relations leads to
\begin{equation}
    \alpha_j = \mbraket{w_j}{T}{v_j} = A_j^{(1)}.
\end{equation}
Similarly combining the recursion relations with $\braket{w_{j+1}}{v_{j-1}} = \braket{w_{j-1}}{v_{j+1}} = 0$ gives
\begin{equation}
\begin{split}
\beta_j &= \mbraket{w_{j-1}}{T}{v_j} =  B_j^{(1)}, \\
    \gamma_j &=  \mbraket{w_{j}}{T}{v_{j-1}} = G_j^{(1)} .
    \end{split}
\end{equation}
Applying the recursion relations to $\braket{w_j}{v_j}$ further gives the relations
\begin{equation}
\begin{split}
    \beta_j = \tau_j,  \hspace{20pt} \gamma_j = \rho_j ,
    \end{split}
\end{equation}
and therefore the inner product of the left- and right-Lanczos vector residuals is
\begin{equation}
    \braket{s_j}{r_j} =  \rho_j \tau_j  = \beta_j \gamma_j.
\end{equation}
These results allow Eq.~\eqref{eq:sr} to be simplified as
\begin{equation}\label{eq:sr2}
    \begin{split}
      \braket{s_{j+1}}{r_{j+1}}  &= A_j^{(2)}  - \alpha_j^2  - \beta_j\gamma_j,
    \end{split}
\end{equation}
which shows how the oblique Lanczos formula for $\beta_{j+1} \gamma_{j+1}$ is equivalent to the symmetric Lanczos formula for $\beta_{j+1}^2$ when $\gamma_j$ is equal to $\beta_j$.
This means that if $\bra{w_1}$ is equal to $\bra{v_1}$, then oblique and symmetric Lanczos results are identical if and only if $\braket{s_{j}}{r_{j}} > 0$ for all iterations. 

\subsubsection{Comparison with symmetric Lanczos}

For a Hermitian operator $T$, all of the elements of $\Tm_{ij}$ defined above are real but are not necessarily positive.
In particular, for steps where $\beta_j \gamma_j < 0$, the above definitions give $\gamma_j = -\beta_j$.
For all steps where $\beta_j \gamma_j > 0$, the results for $\alpha_j$, $\beta_j$, and $\gamma_j$ from applying oblique Lanczos to a Hermitian operator $T$ are identical to those obtained by applying the symmetric Lanczos algorithm described in the main text.
Note that this statement applies even for $j$ larger than a step $k$ in which $\beta_k \gamma_k < 0$ was realized --- the negative $\tau_k$ leads to a sign flip in $\bra{w_k}$ relative to the symmetric case, but this sign is cancelled in the recursions for $\bra{w_{k+1}}$ by a corresponding sign change in $\gamma_k$.
This means that the only difference between applying symmetric and oblique Lanczos to a Hermitian operator $T$ is that $\Tm_{ij}$ includes $\gamma_j = -\beta_j < 0$ when $\beta_j \gamma_j < 0$, while symmetric Lanczos has a purely imaginary $\beta_j$ on both off-diagonals.
These two matrices can be related by a change of basis and therefore have identical eigenvalues $\lambda_n^{(m)}$.
This implies that, with arbitrary-precision arithmetic, applying symmetric Lanczos with $\beta_j \in \mathbb{C}$ gives identical $E_n^{(m)}$ as applying oblique Lanczos even at finite statistics when fluctuations lead to negative estimates of squared residual norms.

Round-off errors are quite different when the symmetric and oblique Lanczos algorithms are performed using floating-point arithmetic. Using oblique Lanczos, $T_k^{(m)}$ is real, and its eigenvalues are therefore either exactly real (at working precision) or come in complex conjugate pairs with non-zero imaginary parts. Using symmetric Lanczos, $T_k^{(m)}$ is generally a complex symmetric matrix, which can have unpaired eigenvalues that are approximately but not exactly real. Distinguishing these approximately real eigenvalues from pairs of complex conjugate eigenvalues is relatively straightforward for the examples studied here, but it adds additional complications that are unnecessary when using oblique Lanczos.

The convergence of oblique Lanczos for non-Hermitian $T_k^{(m)}$ is governed by an analog of the KPS bound that was derived by Saad in Ref.~\cite{Saad:1982}.
Even for the ground state, the bound depends on the entire spectrum rather than just $E_0$ and $Z_0$ as in the Hermitian case.
Monotonicity and one-sided variational bounds no longer apply for non-Hermitian $T_k^{(m)}$; results can approach from above or below.

\subsubsection{Residual bound: right Ritz vectors}

A bound on the residual norms of Lanczos approximations to $T\ket{v_j}$ and $\bra{w_i}T$ can be obtained straightforwardly for oblique Lanczos~\cite{Saad:1982}.
However, Paige's proof of the eigenvalue-level residual bound~\cite{Paige:1971}, Eq.~\eqref{eq:Rbound}, does not apply for non-Hermitian $T$. 
In the remainder of this section, I demonstrate that Paige's proof can be extended to the situation relevant for LQCD in which $T=T^\dagger$ but 
$T_{ij}^{(m)}$ is non-Hermitian due to statistical noise and/or using distinct left- and right-Lanczos vectors. 

To obtain a formula for computing the residual norm, first note that the eigenvectors $\omega_k^{(m)}$ of $\Tm_{ij}$ appearing in the eigendecomposition
\begin{equation}\label{eq:eigen}
\begin{split}
  \Tm_{ij} &= \sum_k \omega_{ik}^{(m)} \lambda_k^{(m)} (\omega^{-1})_{kj}^{(m)},
    \end{split}
\end{equation}
satisfy the eigenvalue equation
\begin{equation}
    \sum_k \Tm_{ik} \omega^{(m)}_{kj} = \omega^{(m)}_{ij} \lambda_j^{(m)},
\end{equation}
which can be seen by matrix multiplying Eq.~\eqref{eq:eigen} by $\omega^{(m)}_{jl}$ on the right and relabeling indices.
The right Ritz vectors, defined by
\begin{equation}
    \ket{y^{(m)}_k} \equiv \sum_i \ket{v_i} \omega_{ik}^{(m)},
\end{equation}
provide the Hilbert-space vectors corresponding to the eigenvectors of $\Tm_{ij}$.
To make this correspondence precise, define a Hilbert space operator $\Tm$ by
\begin{equation}
    \Tm \ket{v_j} \equiv \sum_i \ket{v_i} \Tm_{ij}.
\end{equation}
The matrix elements of this operator are given by
\begin{equation}
\begin{split}
\mbraket{w_i}{\Tm}{v_j} &= \sum_k \braket{w_i}{v_k} \Tm_{kj} \\
&= \sum_k \delta_{ik} \Tm_{kj} \\
&= \Tm_{ij},
\end{split}
\end{equation}
and so that the matrix elements of $T$ and $\Tm$ between Lanczos vectors are both equal to the tridiagonal matrix elements $\Tm_{ij}$.
The Ritz vectors therefore
satisfy the eigenvalue equation
\begin{equation}\label{eq:Ritz_vecs_tri}
\begin{split}
   \Tm \ket{y^{(m)}_k} &= \sum_{i} \Tm \ket{v_i} \omega_{ik}^{(m)} \\
   &= \sum_{i,j}   \ket{v_j} \Tm_{ji}  \omega_{ik}^{(m)} \\
   &= \sum_j \ket{v_j} \lambda^{(m)}_k \omega_{jk}^{(m)}  \\
   &= \lambda^{(m)}_k \ket{ y^{(m)}_k }.
   \end{split}
\end{equation}

Next, note that the action of $T$ is given by Eq.~\eqref{eq:lr_vecs} and $\ket{r_{j+1}} = \rho_{j+1} \ket{v_{j+1}} = \gamma_{j+1} \ket{v_{j+1}}$ as 
\begin{equation}
  T \ket{v_j} = \alpha_j \ket{v_j} + \beta_j \ket{v_{j-1}} + \gamma_{j+1} \ket{v_{j+1}}.
\end{equation}
The analogous action of the tridiagonal matrix $\Tm$ is given directly from Eq.~\eqref{eq:untri} by 
\begin{equation}
\begin{split}
    \Tm\ket{v_j} &= \sum_i  \ket{v_i} \Tm_{ij} \\
    &= \begin{cases}
      \alpha_j \ket{v_j} + \beta_j \ket{v_{j-1}} + \gamma_{j+1} \ket{v_{j+1}}, & j < m \\
        \alpha_j \ket{v_j} + \beta_j \ket{v_{j-1}}, & j = m,
    \end{cases}
    \end{split}
\end{equation}
which leads to
\begin{equation} \label{eq:error}
  [T - \Tm] \ket{v_j}  = \delta_{jm} \gamma_{m+1} \ket{v_{m+1}},
\end{equation}
where $\ket{v_{m+1}}$ is the Lanczos vector obtained by extending from $m$ to $m+1$ steps.
The action of $T-\Tm$ on Ritz vectors is therefore
\begin{equation}\label{eq:Ritz_error}
\begin{split}
   [T - \Tm] \ket{y^{(m)}_k} &= \sum_{j} [T - \Tm]\ket{v_j} \omega_{jk}^{(m)} \\
   &= \gamma_{m+1} \omega^{(m)}_{mk} \ket{v_{m+1}}.
   \end{split}
\end{equation}
Defining dual right Ritz vectors (which are distinct from the left Ritz vectors introduced below) via the usual Hilbert space adjoint, 
\begin{equation}
    \bra{y^{(m)}_k} \equiv \ket{y^{(m)}_k}^\dagger = \sum_i \bra{v_i} [\omega^{(m)}_{ik}]^*,
\end{equation}
taking the Hermitian conjugate of Eq.~\eqref{eq:Ritz_error} gives
\begin{equation}\label{eq:Ritz_error_dag}
  \bra{y^{(m)}_k} [T - \Tm]^\dagger = \gamma_{m+1} [\omega^{(m)}_{mk}]^* \bra{v_{m+1}},
\end{equation}
since the $\beta_j$ and $\gamma_j$ are all real.

These ingredients can be used to compute the Ritz vector residual norm, defined by
\begin{equation}
    \begin{split}
    R_k^{(m)} &\equiv || [T - \lambda_k^{(m)}] \bigl|y_n^{(m)} \bigr> ||^2 \\
    &= \mbraket{ y_k^{(m)}} {[T - \lambda_k^{(m)}]^\dagger [T - \lambda_k^{(m)}] } { y_k^{(m)}}.
    \end{split}
\end{equation}
Using Eq.~\eqref{eq:Ritz_vecs_tri} and its complex conjugate, the factors of $\lambda_k^{(m)}$ can be replaced by $\Tm$,
\begin{equation}
    R_k^{(m)} 
    = \mbraket{ y_k^{(m)}} {[T - \Tm]^\dagger [T - \Tm] } { y_k^{(m)}}.
\end{equation}
Using Eq.~\eqref{eq:Ritz_error} and Eq.~\eqref{eq:Ritz_error_dag} then gives
\begin{equation}\label{eq:ob_res}
    R_k^{(m)} 
    = \gamma_{m+1}^2 |\omega^{(m)}_{mk}|^2 \braket{v_{m+1}}{v_{m+1}}.
\end{equation}
This is the oblique Lanczos analog of the right-hand-side of Eq.~\eqref{eq:Rbound}.

\subsubsection{Residual bound: spectral representation}

Paige's proof connecting the residual bound to eigenvalue error bounds~\cite{Paige:1971} can now be applied.
The proof requires the assumption that $T$ is Hermitian even though non-Hermitian $\Tm_{ij}$ is allowed. 
Denoting as usual eigenstates of $T$ by $\ket{n}$ and their Hermitian conjugates by $\bra{n}$ with normalization $\braket{n}{n} = 1$, the residual norm has a spectral representation,
\begin{equation}\label{eq:res_spec}
\begin{split}
    R_k^{(m)} &= \mbraket{ y_k^{(m)}} {[T-\lambda_k^{(m)}]^\dagger [T-\lambda_k^{(m)} ]}{ y_k^{(m)}} \\
    &= \mbraket{ y_k^{(m)}} {[T-\lambda_k^{(m)^*}] [T-\lambda_k^{(m)} ]}{ y_k^{(m)}} \\
    &= \sum_n \mbraket{ y_k^{(m)}} { T-\lambda_k^{(m)^*} }{n} \mbraket{n}{ T-\lambda_k^{(m)} } { y_k^{(m)}} \\
    &= \sum_n \mbraket{ y_k^{(m)}} {\lambda_n-\lambda_k^{(m)^*}}{n} \mbraket{n}{\lambda_n-\lambda_k^{(m)}} { y_k^{(m)}} \\
    &= \sum_n |\lambda_k^{(m)} - \lambda_n|^2 \left| Z_{kn}^{(m)} \right|^2,
    \end{split}
\end{equation}
where $Z_{kn}^{(m)} \equiv \bigl< n \bigl| y_k^{(m)} \bigr>$.
Defining
\begin{equation}
  \widetilde{\lambda}_k^{(m)} \equiv \min_{\lambda \in \{\lambda_n\}} |\lambda_k^{(m)} - \lambda|,
\end{equation}
where the minimum is over the discrete set of true eigenvalues, i.e. $\tilde{\lambda}_k^{(m)}$ is the closest true eigenvalue to $\lambda^{(m)}_k$,
an inequality for $R_k^{(m)}$ can be derived from Eq.~\eqref{eq:res_spec} as
\begin{equation}
\begin{split}
  R_k^{(m)} &\geq \sum_n |\lambda_k^{(m)} - \widetilde{\lambda}_k^{(m)} |^2 \left| Z_{kn}^{(m)} \right|^2 \\
    &= |\lambda_k^{(m)} - \widetilde{\lambda}_k^{(m)}|^2 \sum_n  \left| Z_{nk}^{(m)} \right|^2,
    \end{split}
\end{equation}
because $\widetilde{\lambda}_k^{(m)}$ is always nearer or identical to the replaced $\lambda_n$.
The overlap factor sum can be expressed as
\begin{equation}
    \begin{split}
        \sum_n |Z_{kn}|^2 &= \sum_n  \braket{y_k^{(m)}}{n}\braket{n}{y_k^{(m)}} \\
        &= \braket{y_k^{(m)}}{y_k^{(m)}} .
    \end{split}
\end{equation}
which allows $R_k^{(m)}$ to be expressed as
\begin{equation}
\begin{split}
  R_k^{(m)} &\geq |\lambda_k^{(m)} - \widetilde{\lambda}_k^{(m)}|^2 \braket{y_k^{(m)}}{y_k^{(m)}} \\ 
    &= \min_{\lambda \in \{\lambda_n\}} |\lambda_k^{(m)} - \lambda|^2 \braket{y_k^{(m)}}{y_k^{(m)}}.
    \end{split}
\end{equation}

Combining this with Eq.~\eqref{eq:ob_res} provides a residual bound for oblique Lanczos,
\begin{equation}\label{eq:ob_bound}
  \min_{\lambda \in \{\lambda_n\}} |\lambda_k^{(m)} - \lambda|^2 \leq B_k^{(m)} \equiv \gamma_{m+1}^2 \,  |\omega_{mk}^{(m)}|^2 \, V_k^{(m)} , 
\end{equation}
where
\begin{equation}
\begin{split}
    V_{k}^{(m)}& \equiv \frac{\braket{v_{m+1}}{v_{m+1}}}{\braket{y_k^{(m)}}{y_k^{(m)}}} \\
    &= \frac{\braket{v_{m+1}}{v_{m+1}}}{\sum_{ij} [\omega_{ik}^{(m)}]^*  \braket{ v_i }{ v_j } \omega_{jk}^{(m)}}.
    \end{split}
\end{equation}
This bound is summarized in Eq.~\eqref{eq:Rbound}

\subsubsection{Residual bound: left Ritz vectors}

An identical derivation can be performed using the left Ritz vectors
\begin{equation}
    \bra{x_k^{(m)}} \equiv \sum_i (\omega^{-1})_{ki}^{(m)} \bra{w_i},
\end{equation}
which satisfy
\begin{equation}
\begin{split}
    \bra{x_k^{(m)}} \Tm &=  \sum_i (\omega^{-1})_{ki}^{(m)} \bra{w_i} \Tm \\
    &= \sum_{ij} (\omega^{-1})_{ki}^{(m)} T_{ij}^{(m)} \bra{w_j} \\
    &= \sum_{j} \lambda_k^{(m)} (\omega^{-1})_{kj}^{(m)} \bra{w_j} \\
    &= \lambda_k^{(m)} \bra{x_k^{(m)}},
    \end{split}
\end{equation}
and their duals $\bigl| x_k^{(m)} \bigr> \equiv \bigl< x_k^{(m)} \bigr|^\dagger$.
Together with
\begin{equation}
  \bra{w_j} [ T - \Tm ]  = \delta_{jm} \beta_{m+1} \bra{w_{m+1}},
\end{equation}
the spectral representation for $|| \bra{ x_k^{(m)}  } [T - \lambda_k^{(m)}]  ||^2$ analogous to Eq.~\eqref{eq:res_spec} can be used to show that
\begin{equation}\label{eq:ob_bound_W}
    \min_{\lambda \in \{\lambda_n\}} |\lambda_k^{(m)} - \lambda|^2 \leq \beta_{m+1}^2 \,  |(\omega^{-1})_{km}^{(m)}|^2 \, W_k^{(m)} , 
\end{equation}
where
\begin{equation}
\begin{split}
    W_{k}^{(m)}& \equiv \frac{\braket{w_{m+1}}{w_{m+1}}}{\braket{x_k^{(m)}}{x_k^{(m)}}} \\
    &= \frac{\braket{w_{m+1}}{w_{m+1}}}{\sum_{ij} (\omega^{-1})_{ki}^{(m)}  \braket{ w_i }{ w_j } [(\omega^{-1})_{kj}^{(m)}]^*}.
    \end{split}
\end{equation}
When $\bra{w_i} = \ket{v_i}^\dagger$, as for symmetric bosonic correlation functions in the infinite statistics limit, then $\Tm_{ij}$ is symmetric, $\omega^{(m)}_{ij}$ is unitary, and both Eq.~\eqref{eq:ob_bound} and Eq.~\eqref{eq:ob_bound_W} reduce to Eq.~\eqref{eq:Rbound}.
In general, Eq.~\eqref{eq:ob_bound} and Eq.~\eqref{eq:ob_bound_W} provide rigorous two-sided bounds on the eigenvalue error $\min_{\lambda} |\lambda_k^{(m)} - \lambda|^2$ that are valid stochastically at finite statistics.
Both bounds hold simultaneously, so the more constraining one may be taken.

\subsubsection{Residual bound: auxiliary recursion relations}

To compute (stochastic estimators for) these bounds in practice, it remains to obtain formulae for $\braket{v_i}{v_j}$ and $\braket{w_i}{w_j}$.
For $\braket{v_i}{v_j}$ these can be obtained using recursion relations for
\begin{equation}\label{eq:r_MEs}
    \begin{split}
        R_{ij}^{(k)} &\equiv \mbraket{ v_i }{ T^k }{ v_j } = R_{ji}^{(k)}, \\
        C_j^{(k)} &\equiv R_{jj}^{(k)},
    \end{split}
\end{equation}
The relevant recursions can be derived as in the symmetric case by inserting Eqs.~\eqref{eq:lr_vecs}-\eqref{eq:lr_norms} into Eq.~\eqref{eq:r_MEs}.
The recursion relation needed to obtain $R_{i(j+1)}^{(k)}$ for $i \leq j$ is
\begin{equation}
    \begin{split}
        \rho_{j+1}  R_{i(j+1)}^{(k)} &= 
            R_{ij}^{(k+1)} - \alpha_j R_{ij}^{(k)} - \beta_j R_{i(j-1)}^{(k)},
            \end{split}
\end{equation}
and the relation needed for $C_{j+1}^{(k)} = R_{(j+1)(j+1)}^{(k)}$ is
\begin{equation}
    \begin{split}
         \rho_{j+1}^2 C_{j+1}^{(k)}  &= C_j^{(k+2)}  - 2 \alpha_j  C_j^{(k+1)} + \alpha_j^2  C_j^{(k)} + \beta_j^2 C_{j-1}^{(k)}  \\
        &\hspace{20pt}   + 2 \alpha_j  \beta_j  R_{j(j-1)}^{(k)}   - 2 \beta_j  R_{j(j-1)}^{(k+1)}.
    \end{split}
\end{equation}
The analogous set of matrix elements required for $\braket{w_i}{w_j}$ is given by
\begin{equation}\label{eq:l_MEs}
    \begin{split}
        L_{ij}^{(k)} &\equiv \mbraket{ w_i }{ T^k }{ w_j } = L_{ji}^{(k)}, \\
        D_j^{(k)} &\equiv L_{jj}^{(k)}.
    \end{split}
\end{equation}
The recursion relation to obtain $L_{i(j+1)}^{(k)}$ for $i \leq j$ is
\begin{equation}
    \begin{split}
        \tau_{j+1}  L_{i(j+1)}^{(k)} &= 
            L_{ij}^{(k+1)} - \alpha_j L_{ij}^{(k)} - \gamma_j L_{i(j-1)}^{(k)},
            \end{split}
\end{equation}
and the relation for $D_{j+1}^{(k)} = L_{(j+1)(j+1)}^{(k)}$ is
\begin{equation}
    \begin{split}
         \tau_{j+1}^2 D_{j+1}^{(k)}  &= D_j^{(k+2)}  - 2 \alpha_j  D_j^{(k+1)} + \alpha_j^2  D_j^{(k)} + \gamma_j^2 D_{j-1}^{(k)}  \\
        &\hspace{20pt}   + 2 \alpha_j  \gamma_j  R_{j(j-1)}^{(k)}   - 2 \gamma_j  R_{j(j-1)}^{(k+1)}.
    \end{split}
\end{equation}
The factors of $V_k^{(m)}$ and $W_k^{(m)}$ appearing in the oblique Lanczos residual bounds, Eq.~\eqref{eq:ob_bound} and Eq.~\eqref{eq:ob_bound_W}, can be computed straightforwardly from these recursion results,
\begin{equation}
    \begin{split}
        V^{(m)}_k &= \frac{C^{(0)}_{m+1}}{\sum_{ij} [\omega_{ik}^{(m)}]^* R^{(0)}_{ij} \omega_{jk}^{(m)}}, \\
        W^{(m)}_k &= \frac{D^{(0)}_{m+1}}{\sum_{ij} (\omega^{-1})_{ki}^{(m)}  L^{(0)}_{ij} [(\omega^{-1})_{kj}^{(m)}]^*}.
    \end{split}
\end{equation}

\subsection{Additional numerical results}\label{sec:plots}

\begin{figure}[t!]
                \centering
                \includegraphics[width=0.48\textwidth]{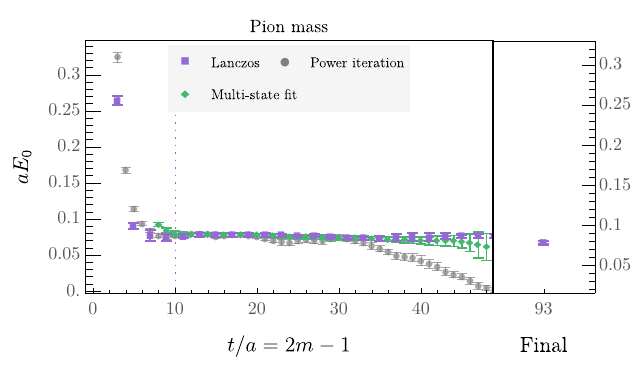}
            \caption{\label{fig:pion_correlator}
            $E(t)$, $E_0^{(m)}$, and multi-state fit results for the pion; details are as in Fig.~\ref{fig:mock_correlator}.
                }
\end{figure}

Pion correlation functions in LQCD have the distinction of not suffering from exponential SNR degradation. Lanczos results for the pion mass are shown in comparison with $E(t)$ and  multi-state fit results in Fig.~\ref{fig:pion_correlator}.
After $m_{\rm max} = 47$ iterations,
$E_0^{(m_{\rm max})} = 0.0779(21)$.
The two-sided window provided by the residual bound $B_0^{(m_{\rm max})}$ is $[0.0612(63), 0.0932(65)]$.

Two-state correlation function fits with $t/a = 7$ give $0.0781(8)$  with $\chi^2/\text{dof} = 0.16$. A weighted average of multi-state fits gives $0.0783(18)$.
A much higher-statistics calculation using 1024 Gaussian-smeared sources on 2511 gauge-field configurations with an identical action and lattice volume finds $am_\pi^{\rm big} = 0.0779(5)$~\cite{Hackett:2023nkr}; nearly identical central values are obtained in Refs.~\cite{Mondal:2020ela,Abbott:2024vhj} while  Ref.~\cite{Yoon:2016jzj} obtained $0.0766(9)$.
Lanczos, two-state, and model-averaged fit results all agree with $am_\pi^{\rm big}$ within $1\sigma$.

\begin{figure}[t]
    \centering
    \includegraphics[width=0.45\textwidth]{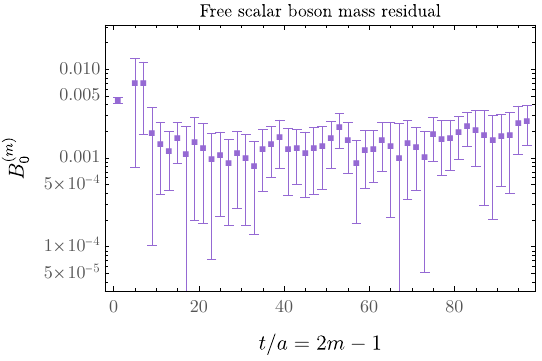} 
    \includegraphics[width=0.45\textwidth]{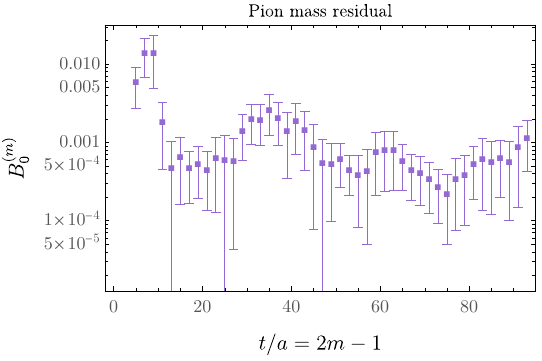} \\
     \includegraphics[width=0.45\textwidth]{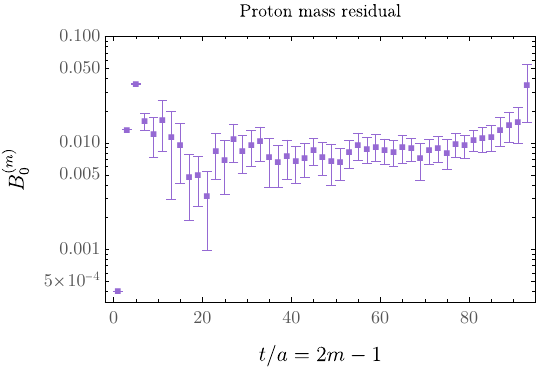}
            \caption{\label{fig:residuals}
            Residual bounds $B_0^{(m)}$, defined in Eq.~\eqref{eq:ob_bound}, for the complex scalar field theory free boson mass, LQCD pion mass, and LQCD proton mass.
            }
\end{figure}

The residual bounds $B_0^{(m)}$ are shown for complex scalar field theory, pion, and proton results in Fig.~\ref{fig:residuals}.

\begin{figure*}[t]
                \centering
                \includegraphics[width=0.32\textwidth]{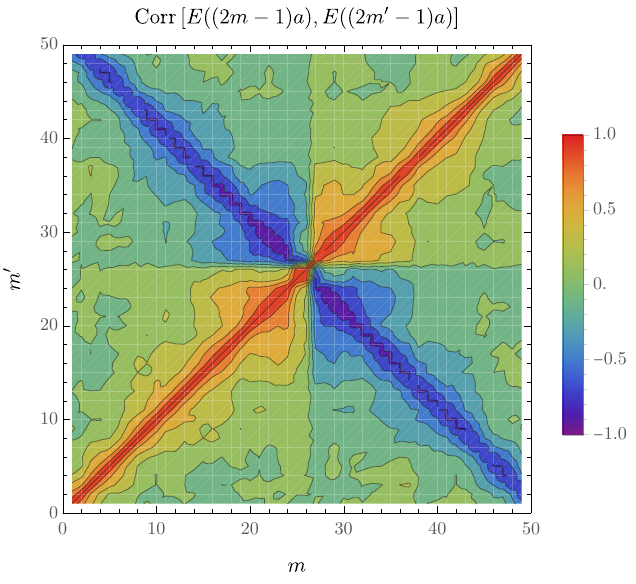} 
                \includegraphics[width=0.32\textwidth]{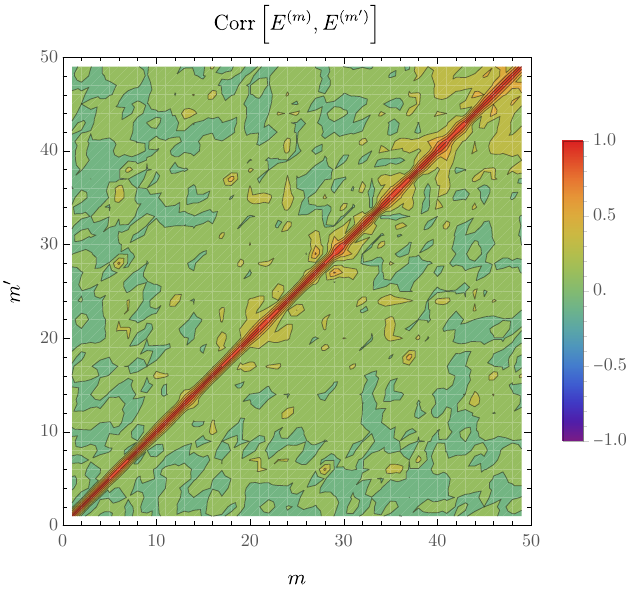} 
                \includegraphics[width=0.32\textwidth]{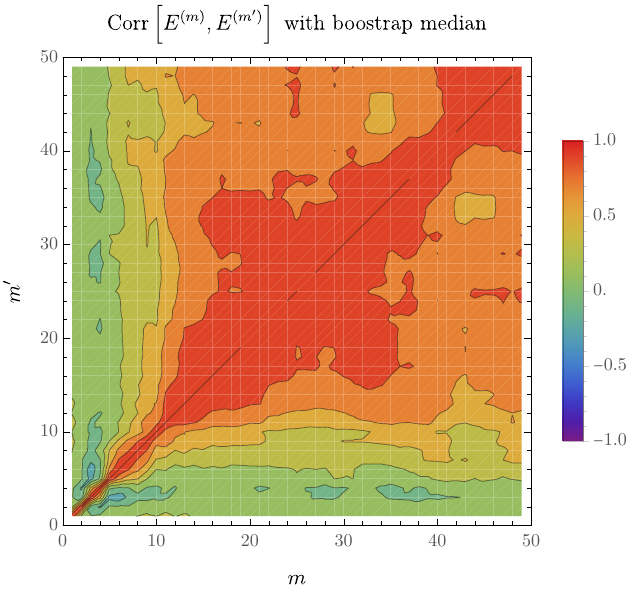} \\
                \includegraphics[width=0.32\textwidth]{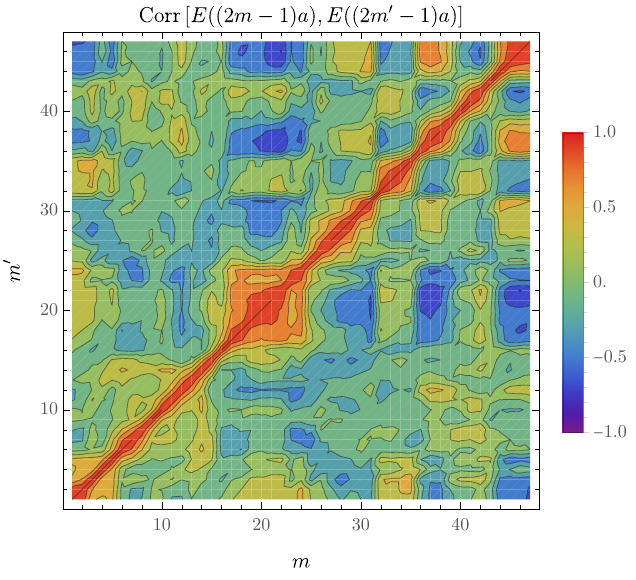} 
                \includegraphics[width=0.32\textwidth]{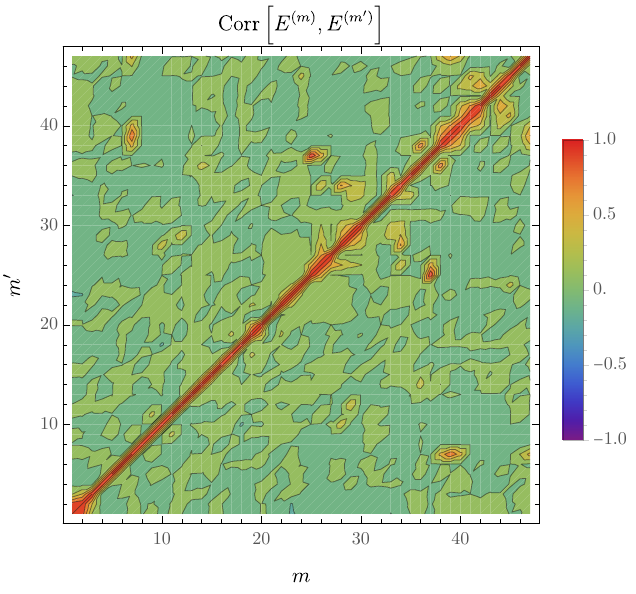} 
                \includegraphics[width=0.32\textwidth]{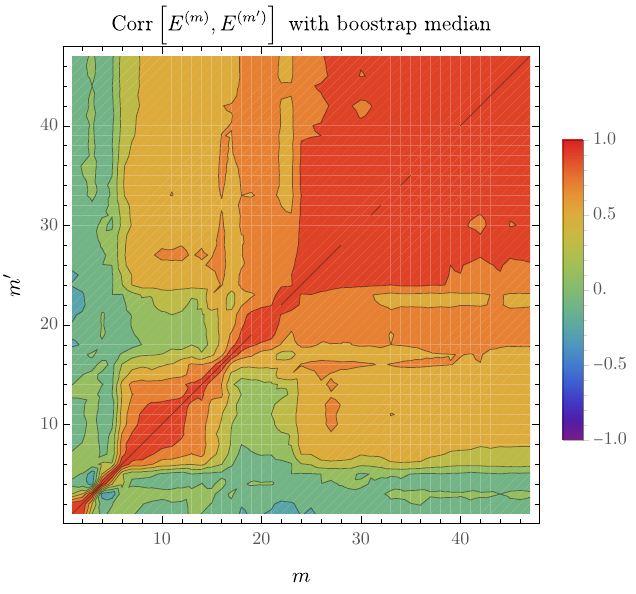} \\
                \includegraphics[width=0.32\textwidth]{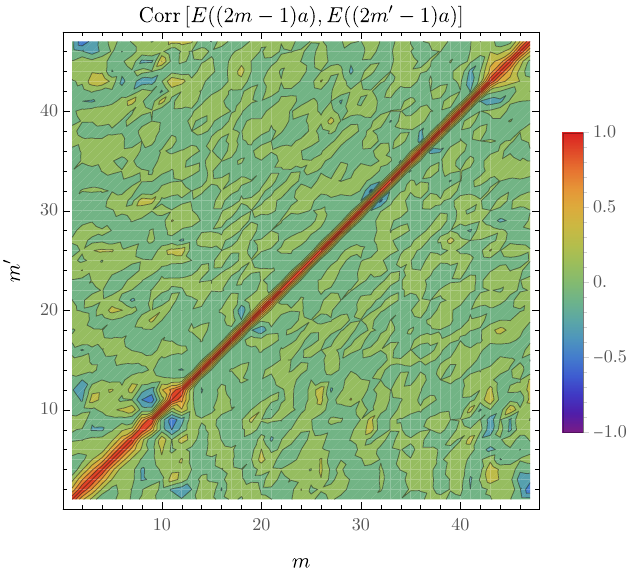} 
                \includegraphics[width=0.32\textwidth]{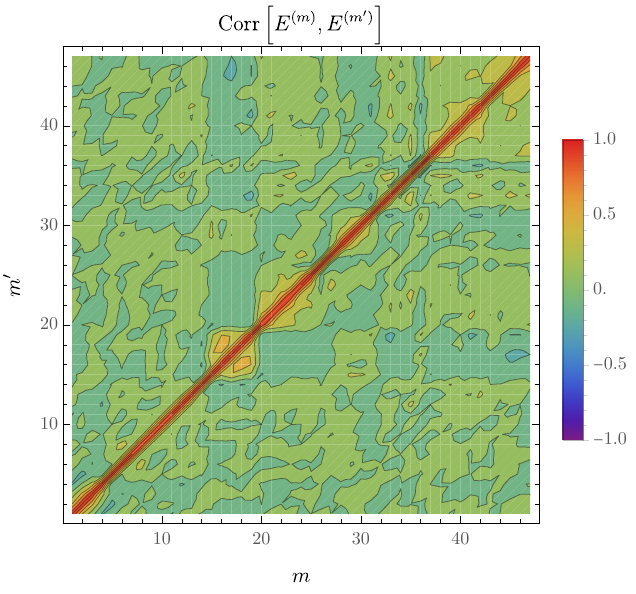} 
                \includegraphics[width=0.32\textwidth]{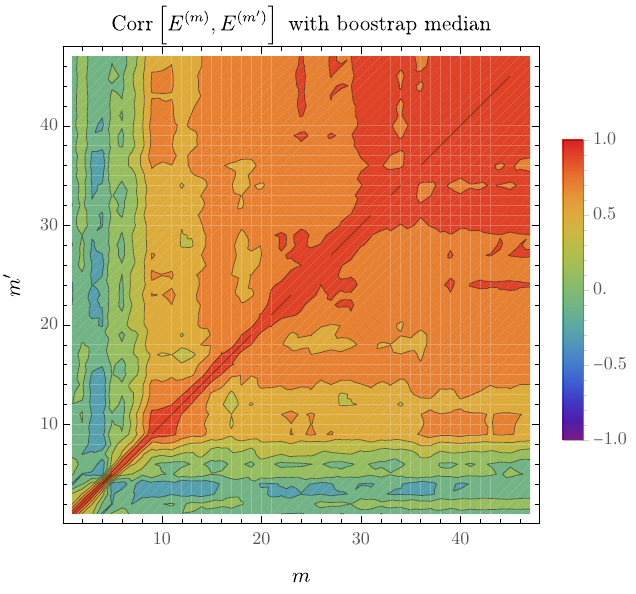} \\
            \caption{\label{fig:cov}
            Correlation matrices for effective mass (left), Lanczos sample mean (middle), and Lanczos bootstrap median (right)  results for complex scalar field theory (top), the LQCD pion mass (middle), and the LQCD proton mass (bottom).
            }
\end{figure*}

Correlation matrices $\text{Corr}[E_0^{(m)}, E_0^{(m')}] \equiv \text{Cov}[E_0^{(m)},E_0^{(m')}]/\sqrt{\text{Var}[E_0^{(m)}]\text{Var}[E_0^{(m')}]}$ are shown for SHO, pion, and proton Lanczos results in Fig.~\ref{fig:cov}.
Analogous effective mass correlation matrices $\text{Corr}[E(t), E(t')]$ with $t = 2m-1$ and $t' = 2m'-1$ are shown for comparison.
In both cases, correlations fall off rapidly away from the diagonal $m=m'$ for all cases studied here.
For the SHO, Lanczos correlations decrease faster with $|m-m'|$ than  effective mass correlations. The standard effective mass also shows significant anticorrelations associated with thermal effects that are absent from Lanczos results.  For the pion, Lanczos correlations  using sample-mean estimators decrease slightly faster with $|m-m'|$ than standard effective mass correlations, while for the proton Lanczos  using sample-mean estimators and standard effective mass correlations show roughly similar decreases with $|m-m'|$.
Bootstrap median Lanczos results show qualitatively different behavior. For small $m$ or $m'$, correlations appear similar to effective mass or sample-mean Lanczos results.
Conversely, for large $m$ and $m'$ correlations approach $O(1)$ values indicating that there is little to no new statistical information gain by increasing $m$.
This saturation should not be surprising if Lanczos is converging to a definite result and there is limited statistical information in relatively imprecise $C(t)$ results with very large $t$.
It is noteworthy that this saturation appears only in bootstrap median results---with sample mean estimators there are additional fluctuations that increase the variance by about an order of magnitude in all of these examples and wash out the correlations between estimators with large $m$ and $m'$.

Bootstrap median and sample mean estimators behave very similarly for small $m$ where few or no spurious eigenvalues are present.
In particular, bootstrap median and sample mean estimators have identical variance when $m=1$ and Lanczos reduces to the effective mass.
Bootstrap median and sample mean estimators show significant differences only when spurious eigenvalues are present.
This suggests that the lack of saturation of correlations seen without bootstrap median is an artifact of spurious eigenvalue misidentification adding large uncorrelated noise to sample mean Lanczos results with large $m$.

\end{document}